\documentclass[pre,twocolumn,longbibliography]{revtex4-1}

\usepackage{bm}
\usepackage{color}
\usepackage{graphicx}
\usepackage{amssymb,amsfonts,mathtools,amsmath,empheq}
\usepackage{afterpage}
\usepackage[titletoc, title]{appendix}
\usepackage{float}

\begin{document}
\title{Towards a theory of assembly of protein complexes: lessons 
from equilibrium statistical physics}

\author{Pablo Sartori}
\affiliation{The Simons Center for Systems Biology,  School of Natural Sciences, Institute for Advanced Study, Einstein Drive, Princeton, NJ 08540, U.S.A.}
\affiliation{Center for Studies in Physics and Biology and Laboratory of Living Matter, The Rockefeller University, 1230 York Ave, New York, NY 01065, U.S.A.}

\author{Stanislas Leibler}
\affiliation{The Simons Center for Systems Biology,  School of Natural Sciences, Institute for Advanced Study, Einstein Drive, Princeton, NJ 08540, U.S.A.}
\affiliation{Center for Studies in Physics and Biology and Laboratory of Living Matter, The Rockefeller University, 1230 York Ave, New York, NY 01065, U.S.A.}

\date{\today}

\begin{abstract}
Cellular functions are established through biological evolution, but are constrained by the laws of physics. For instance, the physics of protein folding limits the  lengths of cellular polypeptide chains. Consequently, many cellular functions are carried out not by long, isolated proteins, but rather by multi-protein complexes. Protein complexes themselves do not escape physical constraints, one of the most important being the difficulty to assemble reliably in the presence of  cellular noise. In order to lay the foundation for a theory of reliable protein complex assembly, we study here an equilibrium thermodynamic model of self-assembly that exhibits four distinct assembly behaviors: diluted protein solution, liquid mixture, ``chimeric assembly'' and ``multifarious assembly''. In the latter regime, different protein complexes can coexist without forming erroneous chimeric structures. We show that two conditions have to be fulfilled to attain this regime: ($i$) the composition of the complexes needs to be sufficiently heterogeneous, and ($ii$) the use of the set of components by the complexes has to be sparse. Our analysis of publicly available databases of protein complexes indicates that cellular protein systems might have indeed evolved so to satisfy both of these conditions.
\end{abstract}

\maketitle

\section*{Introduction}
Protein complexes are assembled with high compositional accuracy, evidenced e.g. by the possibility of crystalization of complexes as large as the ribosome \cite{ban2000complete}. This is remarkable, because during assembly a growing complex has to discriminate its specific components from a multicomponent mixture of hundreds of different protein species that are part of the proteome.  Failure to solve this discriminatory task could result in assembly of chimeric structures composed of fragments from different complexes, impairing normal cellular function \cite{garcia2018infinite}.

Assembly of protein complexes can also be viewed as a second stage of creating functional cellular structures, the first being the assemblage of amino acids into proteins, achieved by ribosomes. A modest alphabet of twenty amino acids encodes thousands of different proteins. Proteins typically contain all twenty amino acids, so that the amino acid usage by proteins is ``dense'' rather than ``sparse''. Nature, furthermore, reuses amino acids many times within the same protein, which makes the compositional heterogeneity of each protein low. This can be contrasted with the assembly of complexes, which seem to use proteins sparsely, so that each complex contains only a small fraction of the available proteome. At the same time, complexes are often highly heterogeneous, i.e. composed of many different protein species \cite{reuveni2017ribosomes}.

\begin{figure}
\centering
\includegraphics[width=0.5\textwidth]{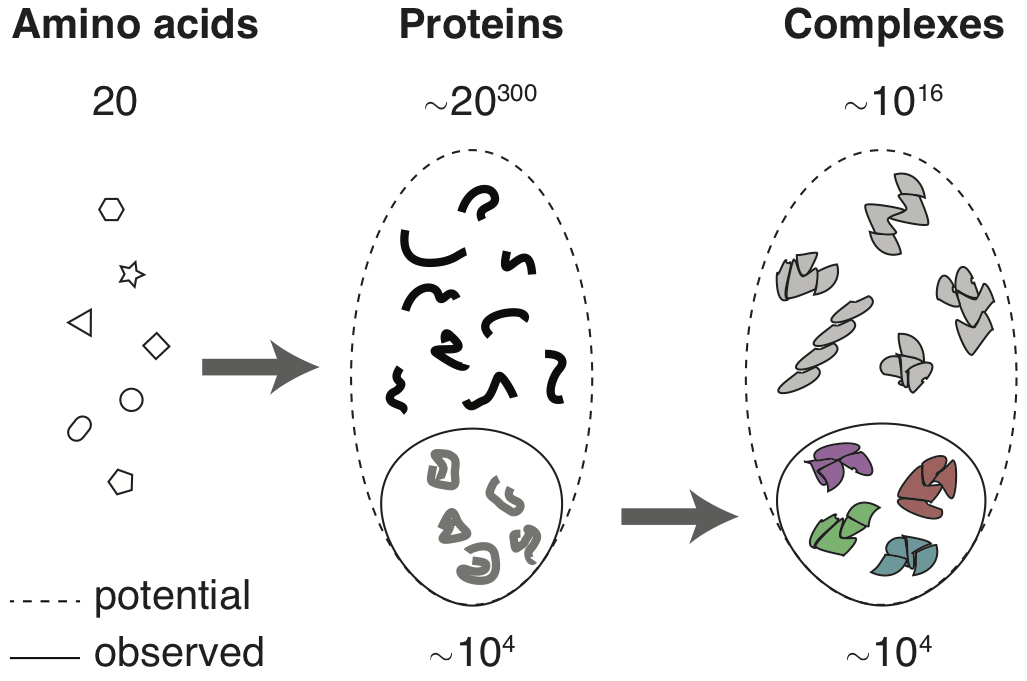}
\caption{
{\bf Usage of amino acids by proteins, and of proteins by complexes.} The typical length of proteins ($\sim 300$ residues) is largely limited by folding \cite{garbuzynskiy2013golden}. 
Twenty amino acids (geometric shapes) potentially thus encode $\sim20^{300}$ proteins (black lines). Although the observed repertoire of proteins (gray lines) is much smaller, e.g. $\sim10^{4}$  for {\it S. cerevisiae} \cite{milo2015cell}, there is a clear combinatorial expansion from amino acids to proteins. Contrary to this, the observed number of complexes, with the reported average of four different proteins per complex \cite{giurgiu2018corum} (colored shapes), is comparable to the number of proteins, without a trace of combinatorial expansion.}
\label{fig:scheme}
\end{figure}

The {\it sparsity} and {\it heterogeneity} of complexes should come as a surprise, as they imply that the proteome might not be exploited in combinatorial manner. Indeed, the vast repertoire of hundreds of proteins is combined to result in a comparable number of complexes, see Fig.~\ref{fig:scheme}. This suggests that ``combinatorial expansion'' of proteins into complexes does not occur generically, and may instead be restricted to particular functions, such as regulation or signaling \cite{stein2009dynamic,  antebi2017combinatorial}. In these cases, proteins participate in several complexes, e.g. cyclin-dependent kinases can be part of several cell cycle regulatory complexes  \cite{murray2004recycling}. Proteins can have specific interactions with many partners, a phenomenon known as {\it promiscuity}. The promiscuity of proteins may potentially result in the formation of disordered chimeric structures. For example, a single point mutation is sufficient to create a novel protein-protein interaction, which can result in chimeric assembly of proteins \cite{garcia2017proteins}. Notwithstanding these challenges, protein complexes typically assemble from their constituents accurately and carry out cellular functions with remarkable speed and precision \cite{johansson2012genetic}.

Elucidating the characteristics of protein complexes that enable them to assemble reliably, and studying how these characteristics affect the organization of the proteome, can be viewed as fundamental goals of cell biology. Recently, there have been significant advances towards achieving these goals due to the progress in experiments \cite{mulder2010visualizing, gavin2006proteome, garcia2017proteins}, bioinformatics \cite{levy20063d, ahnert2015principles} and molecular dynamics simulations \cite{perilla2015molecular}. However, a general theoretical framework to understand protein complex formation and usage is still lacking. One major difficulty in developing such a framework is the large diversity of cellular protein complexes. Some complexes, such as microtubules, exhibit unbounded growth \cite{howard2001mechanics}. Others, such as ribosomes, have a well-defined finite size \cite{ban2000complete}. To complicate matters further, the latter complexes can be further divided among those that exhibit strong symmetries, such as the bacterial flagellar motor \cite{berg2003rotary}, and those that are fully asymmetric, such as ribosomes \cite{ban2000complete}. Whereas the principles of assembly of many symmetric complexes have been studied \cite{ahnert2015principles}, the same is not true for asymmetric complexes.

The aim of this article is to begin to develop a theoretical framework, which could ultimately be applied to (asymmetric) protein complexes, by extending a recent model of self-assembly \cite{murugan2015multifarious}. As it will be emphasized in the Discussion section below, cellular assembly is typically a highly controlled, non-equilibrium kinetic process. Still we will constrain our present theoretical study to equilibrium thermal physics alone and explore what constraints thermodynamics imposes on assembly of complexes. Interestingly, we shall see that these constraints alone can --at least partly-- explain the observed heterogeneity of asymmetric complexes and their sparse usage of the proteome. We will also analyze existing structural, compositional, and interaction data of protein complexes to further evaluate some biological implications of our theoretical findings.

\section*{Results}
\subsection*{Multifarious mixtures of components exhibit four assembly regimes}

In our model, detailed in {\it Materials \& Methods}, protein-like components form a multicomponent mixture. Two components forming part of a same complex interact with binding energy $E$ when they are in close proximity. Conversely, we assume that components not forming part of a same complex have a null binding energy. Such components still can interact non-specifically, provided their concentration, $p$, is large. This model has been formulated and studied previously in Ref.~\cite{murugan2015multifarious}. We extend its analysis to allow for variable heterogeneity and sparsity.

\begin{figure*}
\centering
\includegraphics[width=\textwidth]{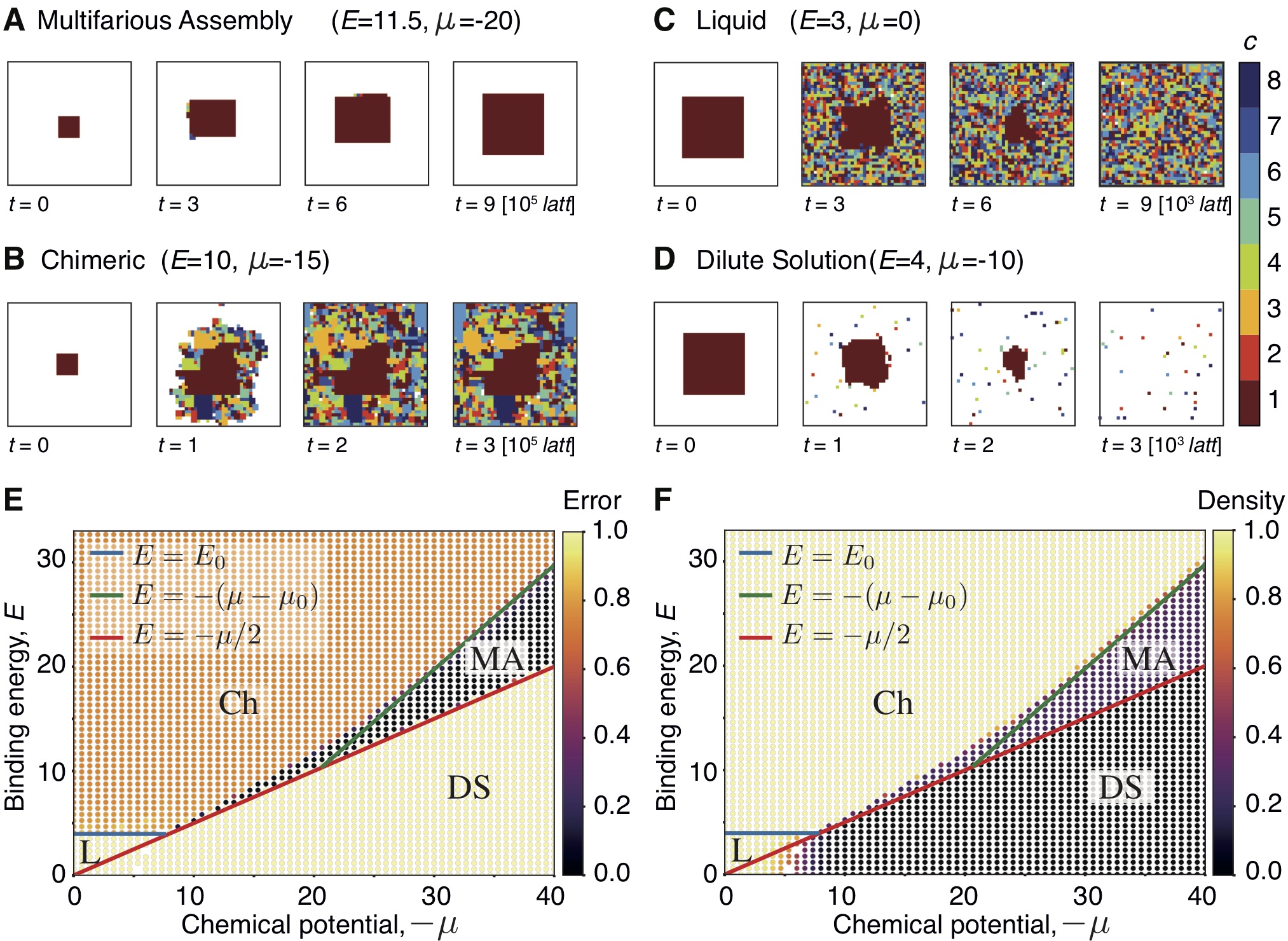}
\caption{
{\bf Four regimes of a multicomponent mixture.} In our simplified equilibrium statistical mechanics model individual different components, $N_{\rm tot}=400$ (small, unit-cell sized, squares), have specific interactions with their neighbors and can form $c=8$ different complexes (large squares) of size $M_c=400$, each represented by a different color (see inset color bar).
{\bf A.} In the multifarious assembly (MA) regime a small fragment of a complex  (small brown square at $t=0$) can be used to nucleate the assembly of the whole complex ($c=1$, large brown square).
{\bf B.} In the chimeric (Ch) regime the same small fragment nucleates a disordered aggregate from fragments of many complexes.
{\bf C.} In the liquid (L) regime, a whole well-assembled complex will be unstable and will ``melt'' into a dense fluid-like mixture, in which small fragments of complexes are constantly being rearranged (unlike in Ch, where there are no rapid rearrangement dynamics).
{\bf D.} In the dilute solution (DS) regime the initial full complex dissolves quickly into a set of separate proteins. Only very small transient clusters of proteins can form.
{\bf E} and {\bf F.} The four regimes observed in A-D correspond to distinct values of the chemical potential and binding energy, and can be determined by the assembly error (portion of the correctly assembled complex) and the density of components. In {\it Supplementary Information} ({\it SI B}) we characterize the boundaries that separate the regimes. Note that each  circle in these graphs correspond to separate Monte Carlo simulations, in which the error of assembly is evaluated. See {\it Materials \& Methods} for detailed description of the parameters used in this and other figures.
}
\label{fig:phases}
\end{figure*}

Just like changing the temperature and pressure of a gas can turn it into a liquid, changing the chemical potential $\mu=\log(p)$ and the binding energy $E$ of the component mixture can fundamentally alter its properties (here energy is expressed in units of $k_{\rm B}T$, with $T$ the temperature and $k_{\rm B}$ Boltzmann's constant). As shown in Fig.~\ref{fig:phases} (which describes a lattice implementation of our model, see also {\it Supplementary Information A}, {\it SI A}), for low negative values of the chemical potential the mixture is in a dilute solution (DS) regime, in which components interact only transiently with each other. This is the regime in which biochemical reactions have been traditionally studied \cite{hill2004free}. If, on the other hand, the chemical potential is high but the binding energy is low, the mixture increases its density and behaves as a liquid (L). The properties of this liquid are somewhat similar to those of multi-component droplets that have recently gained prominence in cell biology  \cite{brangwynne2009germline, sear2003instabilities}. Finally, if  both the chemical potential and the binding energy are high, the liquid mixture changes into a ``chimeric'' (Ch) regime, in which fragments of several protein complexes bind unruly to each other. This regime can evoke cellular inclusion bodies, where over-expressed recombinant proteins form disordered solid aggregates \cite{kopito2000aggresomes}. These three regimes are conceptually close to phases of inert materials, and will not be discussed here further (see, however,  {\it SI A--D}).

We shall rather focus our attention on a more biologically interesting regime, which arises when the values of binding energy and chemical potential are comparable. In this ``multifarious assembly'' (MA) regime a large protein complex can self-assemble accurately, e.g. starting (nucleating) from a small part of such complex (nucleation seed), see Fig.~\ref{fig:phases} \cite{murugan2015multifarious}. One fundamental factor that determines the boundaries of the MA regime is the binding energy, $E$, used to discriminate between specific and non-specific interactions. We expect the range of concentrations $p_{\rm max}/ p_{\rm min}$, in which reliable assembly is possible, to scale exponentially with $E$ (i.e., chemical potentials scale linearly with the energy). For the model at hand, we find indeed that (see {\it SI B}): 
\begin{align}\label{eq:cmaxmin}
p_{\rm max}/ p_{\rm min}\approx\exp\left(E+\mu_0\right)\quad,
\end{align}
see also Figs.~\ref{fig:phases}E and F. Eq.~\ref{eq:cmaxmin} implies that a binding energy of a few $k_{\rm B}T$ is sufficient to ensure reliable assembly in a range of concentrations spanning several orders of magnitude. The parameter $\mu_0<0$ depends logarithmically on characteristics of the mixture: the number of component types, or ``species'', $N_{\rm tot}$, and the number of different complexes, $K$. It also depends on characteristics of a typical complex $c$: the total number of components it contains (i.e. its size), $M_c$, the number of different species among these components, $N_c$, and the number of specific bonds per protein, $z_c$. For the sake of simplicity, we will limit ourselves in the following to the case $z_c=4$, i.e. to a square lattice. However, our results can be generalized to other values of $z_c$, see {\it SI C}.

\subsection*{Heterogeneity and sparsity constrain reliable assembly}

The accurate assembly of complexes is a daunting discriminatory task. Proteins accomplish this task because their interactions and the composition of complexes they form are the result of ``constrained evolution''. That is, the characteristics of complexes have evolved to ensure their cellular function, while at the same time they have been constrained to assemble reliably. As argued above, an important quantity, which is closely related to the reliability of the assembly process, is protein promiscuity: if the promiscuity of a protein were exceedingly high, undesired protein species could interfere with this protein during the assembly process, which would then result in the formation of ``chimeric structures''. Therefore, we expect that evolution tuned protein promiscuity so that proteins do not form such non-functional chimeras.

\begin{figure}
\centering
\includegraphics{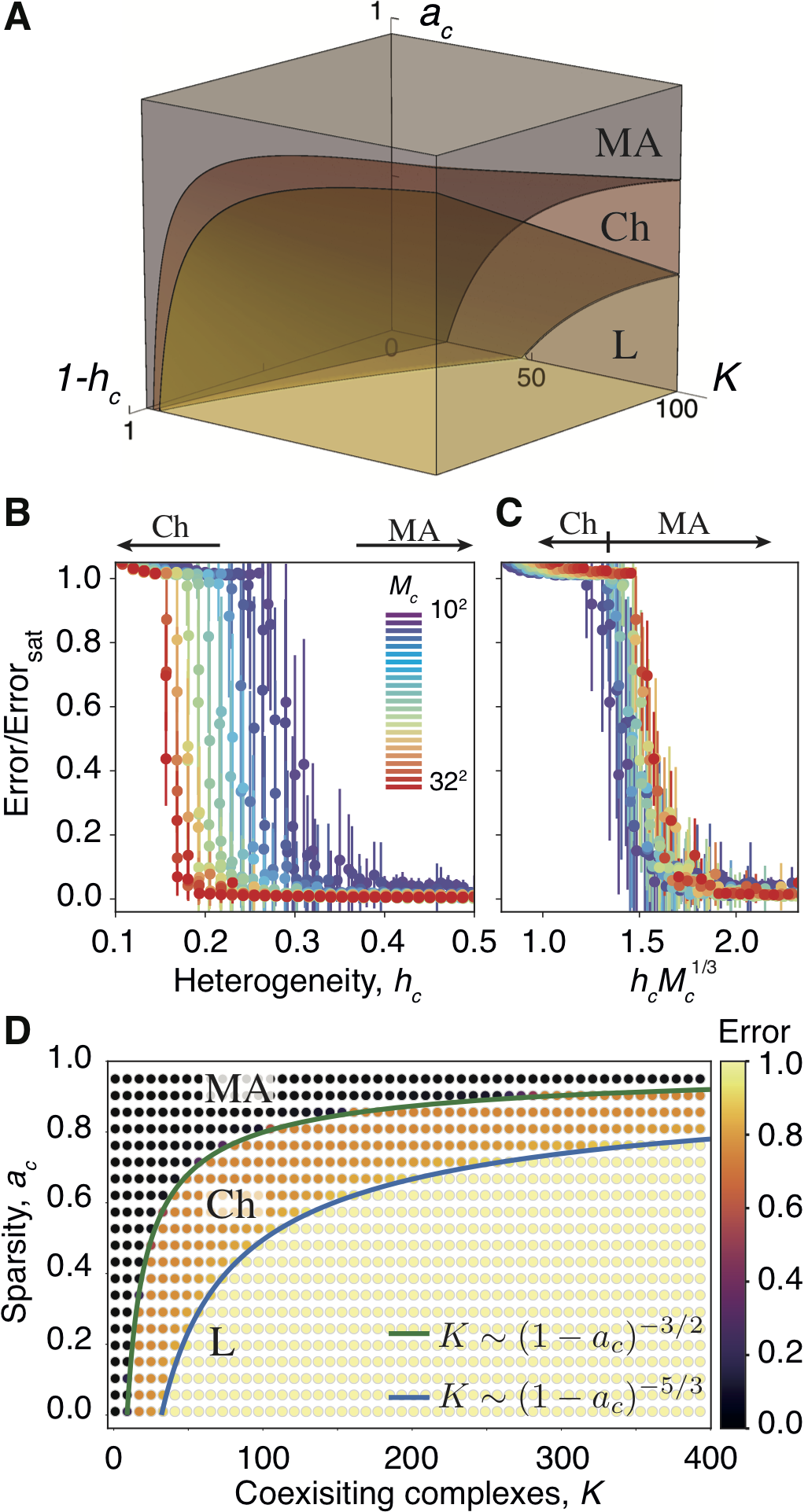}
\caption{
{\bf Characteristics of complexes shape the multifarious assembly regime.}
{\bf A}. Schematic representation of how the different regimes depend on the number of complexes $K$, their heterogeneity $h_c$, and their sparsity $a_c$. Sparsity greatly increases the number of complexes that can be reliably assembled by expanding MA.
{\bf B.} As the heterogeneity of a complex increases, its assembly error decreases (normalized by a saturation error). The system thus transitions from Ch to MA. 
{\bf C.} Data collapse of the data in B using $h_c M^{1/3}_c$, see also {\it SI D}.
{\bf D}. Cross-section of the phase-diagram sketched in A. As the sparsity increases, the number of coexisting complexes diverges algebraically.
}
\label{fig:het_sparse}
\end{figure}

In our model the promiscuity of a protein-like component is related to the number of different complexes, $K$, and to their characteristics, such as their compositional heterogeneity, $h_c\equiv N_c/M_c$; and their sparsity, $a_c\equiv(N_{\rm tot}-N_c)/N_{\rm tot}$, i.e. the fraction of their proteome usage. One can show that the promiscuity of a component species $\alpha$ scales as:
\begin{align}\label{eq:pa}
\pi_\alpha\sim K(1-a_c)/h_c\quad,
\end{align}
(see {\it SI C} for a detailed derivation). One can then express the evolutionary constraint of reliable self-assembly by relating the probability of the formation of chimeric structures to component promiscuity, and requiring that this probability is negligible. By doing so, we find that the number of possible coexisting complexes, their heterogeneity, their size, and their sparsity obey an important constraint relation:
\begin{align}\label{eq:bound1}
K\lesssim  h_c^{3/2}M_c^{1/2}(1-a_c)^{-3/2}\quad,
\end{align}
where the values of the exponents are given for $z_c=4$, see {\it SI C} for generalization. Eq.~\ref{eq:bound1} provides the scaling for the surface of transition between the MA regime, in which chimeric structures are avoided, and the Ch regime, in which they readily form (depicted in Fig.~\ref{fig:het_sparse}A). Although this relation holds in the limit $M_c\to\infty$, our analysis of Monte-Carlo simulations is compatible with Eq.~\ref{eq:bound1} (however, to unequivocally determine the scaling would require a much larger range of complex sizes, see {\it SI D}).

The key determinants of the transition from the MA regime to the Ch regime, predicted by the constraint of Eq.~\ref{eq:bound1}, might be biologically relevant. This constraint can be understood as a trade-off between the number of coexisting complexes, their heterogeneity, and their sparsity: more complexes can coexist if they are more heterogeneous, and if they make a sparser usage of the proteome. To see that this is indeed the case, let us first consider a single complex type ($K=1$) prepared in a mixture that contains only its constituent components, so that $N_{\rm tot}=N_c$ and $a_c=0$. Eq.~\ref{eq:bound1} constrains the heterogeneity of the complex to be larger than a quantity that scales as a power of its size, $h_c\gtrsim M^{-1/3}_c$. Therefore, by increasing the heterogeneity of the complexes, one crosses a transition from the chimeric regime to the desired multifarious assembly regime, as shown in Figs.~\ref{fig:het_sparse}B and C. This mechanism might thus explain the observed high heterogeneity among cellular protein complexes as the means of avoiding assembly of incorrect chimeric structures.

Next, let us consider the possibility of combinatorial usage of components in different complexes. In the case of a dense usage of the set of components, i.e. $a_c=0$, the reliable assembly constraint, Eq.~\ref{eq:bound1}, implies that the number of possible coexisting complexes is $K\lesssim h_cN_{\rm tot}^{1/2}$. Such increase of the number of complexes with increasing number $N_{\rm tot}$ of component species implies that combinatorial usage of components in complexes is indeed possible  \cite{murugan2015multifarious}. However, the reliability constraint makes the combinatorial aspect only sub-linear, and therefore weak: an increase by a factor of one hundred in the number of component species merely increases by ten the number of possible complexes. Therefore, from a biological perspective, reliable assembly introduces a constraint that vastly reduces the possibilities of combinatorial expansion from proteins into protein complexes.

In order for many complexes to coexist, an alternative to this weak combinatorial usage is needed. This alternative is achieved by letting complexes make a sparse usage of the set of components, i.e. $a_c\lesssim 1$. To see this, note that the number of possible coexisting complexes, $K$, diverges in Eq.~\ref{eq:bound1} as the component usage becomes more and more sparse, $a_c\to1$, see also Fig.~\ref{fig:het_sparse}D. An important consequence of such behavior is that the number of coexisting complexes scales super-linearly with the number of component species when a sparse usage is allowed,
\begin{align}\label{eq:sparse_scale}
K\sim N_{\rm tot}^{3/2}M_c^{-1}\quad.
\end{align}
From a biological perspective, due to the previously evoked sparsity of the proteome usage, an increase by a factor of one hundred in the number of protein species results thus in an increase by a factor of a thousand in the number of complexes. Therefore, a sparse usage of the proteome may indeed help to insure the observed coexistence of many different types of protein complexes within the cell.

\begin{figure}
\centering
\includegraphics{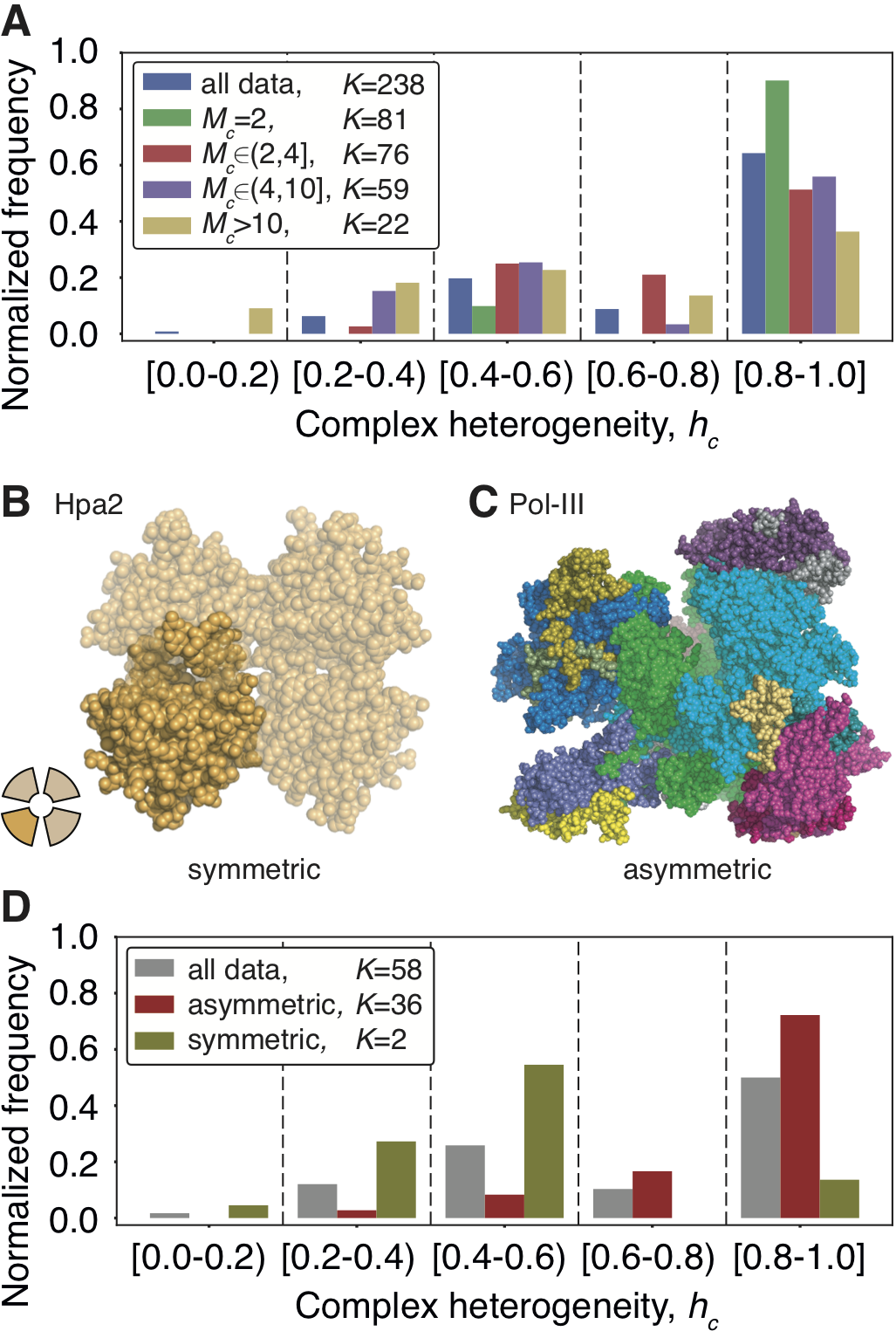}
\caption{
{\bf Heterogeneity of symmetric and asymmetric complexes.}
{\bf A.} Histogram of complex heterogeneity using data from \cite{meldal2018complex}. Most complexes are highly heterogeneous, but a small peak is also present for intermediate values of the heterogeneity. This distribution is preserved across complexes in different size ranges (different colors, see inset legend).
{\bf B} and {\bf C.} Structure of a complex with dihedral symmetry, Hpa2 (PDB: 1QSM), and an asymmetric complex, Pol-III (PDB: 5FJ9). The first has low heterogeneity, $h_c=1/4$, whereas the second is completely heterogeneous, $h_c=1$.
{\bf D.} We cross-referenced data from \cite{meldal2018complex} with structural data taken from the PDB database \cite{berman2000protein}. We then separated symmetric and asymmetric complexes. The high heterogeneity peak is only present for asymmetric complexes, and the peak at intermediate heterogeneities is only present for symmetric complexes. }
\label{fig:dat_het}
\end{figure}

 \subsection*{Structural and compositional data points towards high heterogeneity of protein complexes}

How relevant are the theoretical arguments presented above for cellular protein complexes? It is clear that the proteome usage is generically sparse: even a complex with as many different protein species as the ribosome, for which $N_c\lesssim10^2$, contains only a small fraction of the proteome, $N_{\rm tot}\gtrsim 10^3$. This gives a lower bound for the sparsity of complexes: $a_c>0.9$. However, the second issue of whether highly heterogeneous complexes might coexist more easily, and thus would be more abundant in the cell, could be considered a subject of debate.

To address this question, we analyzed a publicly available database of protein complexes \cite{meldal2018complex}. The resulting histogram of the heterogeneity of these complexes, separated by their size, (see {\it SI H} for details), is depicted in Fig.~\ref{fig:dat_het}A. The histogram reveals a large abundance of high-heterogeneity complexes, $h_c\in[0.8,1.0]$, which supports the arguments presented in this work. From Fig.~\ref{fig:dat_het}A it is also apparent that a significant number of complexes have intermediate values of heterogeneity, $h_c\in[0.4,0.6)$.

Given that our theory applies to structures without any geometrical symmetry (as described in {\it Materials \& Methods}), we can ask whether geometrical symmetry underlies the low heterogeneity of these complexes. For example, a complex with precisely four copies of the same protein, $h_c=1/4$, may be reliably assembled if the proteins are wedged so that they lock in a symmetric ring-like structure, see Fig.~\ref{fig:dat_het}B. The constraints that limit assembly reliability in this type of complexes are likely to be different from those for asymmetric complexes, see Fig.~\ref{fig:dat_het}C. To estimate the role of symmetry in the heterogeneity of complexes, we classified them into those, for which the crystal structure exhibits symmetry, and those, for which it does not \cite{berman2000protein}. In Fig.~\ref{fig:dat_het}D we plot the corresponding heterogeneity histograms, which clearly show that asymmetric complexes have a very large heterogeneity bias, whereas symmetric complexes exhibit a large peak for intermediate heterogeneity values. We thus corroborate that high heterogeneity is indeed widespread among asymmetric complexes, to which we have limited our model. These conclusions should be taken with a grain of salt, though, since generally speaking the databases of protein complexes are in their infancy, and are prone to many possible methodological biases and ambiguities (see also discussion in {\it SI G}).

\section*{Discussion}

\subsection*{Additional mechanisms to prevent chimeric assembly}

Within the cell, assembly of complexes takes place in a dense mixture of proteins. At the basis of successful assembly lays the discrimination of the particular proteins of a complex among many others present in the mixture. Here, we argued that thermal physics puts strong constraints on the characteristics of complexes so that they assemble reliably. In particular, to keep protein promiscuity low, the heterogeneity and sparsity of complexes is constrained to high values. Clearly, within the cell there are additional mechanisms that may help avoid chimeric assembly, some of which we describe now.

First, cellular protein concentrations can be  spatially and temporally controlled to increase assembly precision and yield. In particular, cellular liquid droplets can provide local environments of high concentration of certain proteins, which may increase the assembly yield of corresponding complexes. A well known example of cellular ``compartmentalization'' is the assembly of ribosomes inside the nucleolus, where ribosomal proteins are synthesized \cite{sirri2008nucleolus}. Similarly, the temporal aspects of production and transport of proteins can be regulated so to facilitate and optimize complex assembly \cite{kalir2001ordering}.  Second, it is highly plausible that the energy landscape for protein-protein interactions itself might have evolved so to facilitate the kinetics of protein complex assembly. This is similar to the arguments given for other classical discrimination phenomena in biology, such as the discrimination of correct initiation DNA sites by transcription factors \cite{tafvizi2011dancing, cencini2017energetic}. A third possible mechanism to prevent formation of chimeric structures is the usage of non-pairwise protein interactions, e.g. allostery. For example, it was suggested in Ref.~\cite{antebi2017combinatorial} that different tetrameric receptor complexes of the BMP pathway assemble upon binding of particular ligands. Furthermore, because only seven different species of proteins are involved in forming these receptors, the usage made of these proteins is dense, rather than sparse, which suggests that allostery provides a means to make a dense usage of proteins and enable combinatorial expansion. Finally, the geometry of complexes itself might have also evolved to optimize assembly. The presence of a peak at intermediate values of the heterogeneity in Fig.~\ref{fig:dat_het}A, which can be ascribed to the symmetry of complexes, strongly suggests that the heterogeneity constraint that we derived may be avoided by means of geometric constraints to the structure of the complex, in line with the findings in Refs \cite{ahnert2015principles, garcia2017proteins}.

Two organizational principles of cellular protein assembly explored in our theoretical model, namely the high heterogeneity of protein complexes, and the sparse usage they make of the proteome, should be already functional at thermal equilibrium. So should be also the four additional mechanisms described above; therefore, they could be incorporated and studied within future extensions of our model.  Going beyond equilibrium considerations, it is important to stress that the accuracy and the speed in protein assembly are very likely to be enhanced by a number of out-of-equilibrium mechanisms \cite{hopfield1974kinetic, pigolotti2016protocols, bisker2018nonequilibrium}. For example, more than two hundred non-ribosomal proteins are involved in ribosomal biogenesis \cite{kressler2010driving}. Many of these are energy-consuming enzymes and have a variety of roles, e.g. stabilizing protein-RNA interactions. At the same time, {\it in vitro} studies have shown that it is possible to assemble ribosomal complexes in absence of such enzymes \cite{mizushima1970assembly, rohl1982assembly, mulder2010visualizing}, albeit with a smaller yield. This evokes a possible analogy with the process of protein folding, which is typically facilitated and sped up by energy consuming chaperones, but also can take place in their absence  \cite{hingorani2014comparing}. It is tempting to propose that just as evolution had selected {\it foldable} proteins from the vast space of possible amino acid chain sequences, it might have also selected reliably {\it self-assembling} cellular complexes from the vast space of all possible multi-protein assemblies, see Fig.~\ref{fig:scheme}.

\subsection*{ Broad distribution of protein participation in complexes and dynamic control}

An additional role of out-of-equilibrium processes is to control the dynamics of protein complexes. Indeed, many complexes are assembled in a contextual manner, i.e. only when they carry out a function that is needed. Such highly dynamic complexes are often involved in regulatory and signaling functions \cite{stein2009dynamic, antebi2017combinatorial}, and they can be contrasted with other more stable complexes (such as the ribosome), on which we have focused our attention here. The complexes including cyclins, which participate in the regulation of the cell cycle, provide an example of such highly dynamic complexes \cite{murray2004recycling}. The temporal sequence of assembly and disassembly events observed in this system is correlated with phosphorylation / dephosphorylation of the cell cycle components. Remarkably, the heat release in this out-of-equilibrium process has also been recently measured \cite{rodenfels2019heat}.

\begin{figure}
\centering
\includegraphics{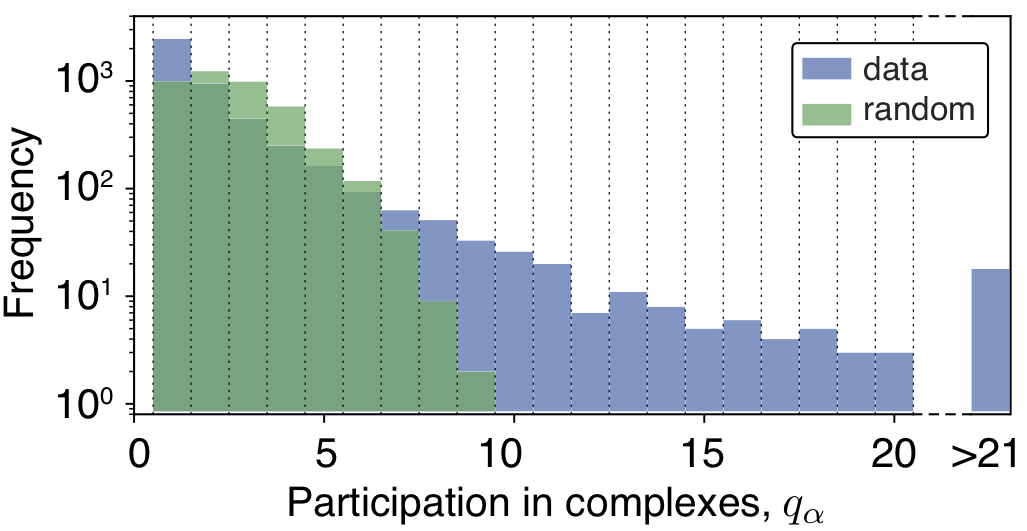}
\caption{
{\bf Protein participation in complexes is not explained by null model.}
In blue, histogram for the participation of proteins in complexes (i.e., in how many complexes a given protein takes part) using the ``core'' dataset from \cite{giurgiu2018corum}, see {\it SI I}. In green, the same histogram for a dataset constructed by randomly grouping protein species into sets with the empirical composition of complexes. The random model largely deviates from the data, and does not account for the prevalence of highly participatory proteins.
}
\label{fig:dat_qa}
\end{figure}

A possible footprint of regulated out-of-equilibrium phenomena is the existence of proteins that participate in many complexes, such as cyclin dependent kinases, which have potential to act as dynamic controllers.  In order to assess the prevalence of such proteins, we analyzed several databases that contained information of protein participation in complexes (see {\it SI I}). Our analysis suggests that the number of complexes, $q_\alpha$, in which a given protein species $\alpha$ participates has a broad distribution, depicted in Fig.~\ref{fig:dat_qa}. Notably, this distribution cannot be simply explained by randomly grouping proteins into sets with the observed composition of complexes, unlike in other systems with shared components \cite{mazzolini2018statistics}. The ``excess'' of highly participatory proteins can be viewed as an indication that this broad distribution might have evolved { to carry out a particular function. One such function could be} to  assure an appropriate dynamic control of different assemblies (provided of course that this broad distribution is confirmed by future analysis of larger databases). A fascinating topic for future research should be, therefore, to extend our theoretical framework so to allow dynamic, out-of-equilibrium control of complexes, and to verify whether the latter could indeed correlate with the observed excess of highly participatory proteins.

\subsection*{ Protein complexes as a distinct regime of matter}
Reliable self-assembly of protein complexes within the noisy cellular environment is intriguing not only from the point of view of cell biology, but also from that of material science. Typical materials contain only a handful of different sorts of atoms, and the possible set of interactions between components is small. This results in states of matter that are easily reproducible: all crystals of salt are formed by Na and Cl atoms arranged in the same way. A different scenario is that of glassy materials, such as silicate glasses, in which the set of effective interactions among constituents is large due to spatial disorder. The large number of interactions makes the state of a glass unique and irreproducible: in each piece of glass the arrangement of atoms is different.

Protein complexes combine similarities with both types of materials. Like glasses, they present a large number of effective interactions, although its origin is not disorder, but the large number of different specifically interacting components. As in crystals, the arrangements of these components may be highly reproducible: all ribosomes in a cell are made of many different proteins, yet the arrangement of the core ribosomal proteins is basically the same. This unusual combination of properties is rarely considered in physical theories of matter, with the closest analogue being ``programmable materials'' of biological origin, such as DNA origami  \cite{jones2015programmable} or self-assembling colloidal particles \cite{zeravcic2017colloquium}. One of the directions of future studies could be to further explore underlying principles of reliable equilibrium and non-equilibrium assembly for synthetic materials inspired by biology.

\section*{Materials and Methods}
\subsection*{Model}

Consider a set of $N_{\rm tot}$ component species labeled $\alpha=1,\ldots ,N_{\rm tot}$, which form the ``proteome'' of our theory. Components can assemble into $K$ different complexes labeled $c=1,\ldots,K$.  How two components of different species, $\alpha$ and $\beta$, interact when they are in proximity is characterized by the binding energy matrix $U_{\alpha\beta}$. If these two components are part of a same complex $c$ in which they are bound to each other, they will interact strongly with an energy $E$, and so $U_{\alpha\beta}=-E$. Conversely, if the components $\alpha$ and $\beta$ are not bound together in any complex, their interaction energy will be assumed null: $U_{\alpha \beta}=0$.  However, even when the interaction energy between two components is null, these may still bind to each other through non-specific interactions, provided that their concentration, $p$, viz. their chemical potential $\mu=\log(p)$, is large enough. Note that we have made the simplifying assumption that the strength of all interactions and the concentrations of all components are the same; this assumption is relaxed in {\it SI F}. One important quantity that characterizes the interactions of a component species $\alpha$ is its {\it promiscuity}, $\pi_\alpha = \sum_\beta \Theta(-U_{\alpha\beta})$, which is the total number of different species with which it has specific interactions. We also define the participation of the species, $q_\alpha$, as the number of different complexes in which it takes part.

Although the structural or enzymatic characteristics of complexes largely define their function, here we are only interested in the characteristics that determine self-assembly. We quantify these characteristics through a small set of parameters.  For each complex $c$, we define its {\it heterogeneity}, $h_c \equiv N_c/M_c$, as the ratio of the number of different component species in the complex, $N_c$, to the total number of components in the complex or complex size, $M_c$ (we have made the assumption that all complexes have the same number of components, relaxed in {\it SI G}). We also define the sparse usage that a complex makes of the available components, i.e. its {\it sparsity}, as $a_c \equiv (N_{\rm tot}-N_c)/N_{\rm tot}$. Finally, we absorb all geometrical properties of a complex into its mean coordination number,  $z_c$, which counts the average number of specific bindings per component in the structure. The latter is an important simplification, which may particularly affect certain complexes, for which the assembly is strongly tied to their geometry \cite{ahnert2015principles}. Note that the model in \cite{murugan2015multifarious} corresponds to the case $h_c=1$ and $a_c=0$ (in addition to fixing the ratio $\mu/E$). Here, we relax these constraints, which allows us to explore the biologically relevant regime of protein complex assembly.

\subsection*{Figure parameters}

In Fig.~\ref{fig:phases} we considered $N_c=M_c=N_{\rm tot}=20^2$ and $K=8$. The initial state  corresponds to a fragment of the complex containing its central components (a nucleation seed). The fragment is of size $7\times7$, in A and B, and $20\times20$ (whole complex) in all other panels. In E and F each point reports the average of $3$ replicate simulations (randomized complexes). Each simulation is run for a duration of $10^6\,{\rm latt}$ in a $40\times40$ lattice, with $1\,{\rm latt}$ corresponding to one lattice sweep. In Fig.~\ref{fig:het_sparse} B and C the parameters are as in Fig.~\ref{fig:phases}, with $E=7$, $\mu=-12.6$, and $K=1$. In D, the parameters are as in B and C, with $h_c=1$ and variable $K$. 

\bibliography{het_spar}

\clearpage
\begin{appendices}
\onecolumngrid
\section*{Theoretical analysis}

\subsection{Computational model}
\label{sec:compu}

\begin{figure}
\centering
\includegraphics{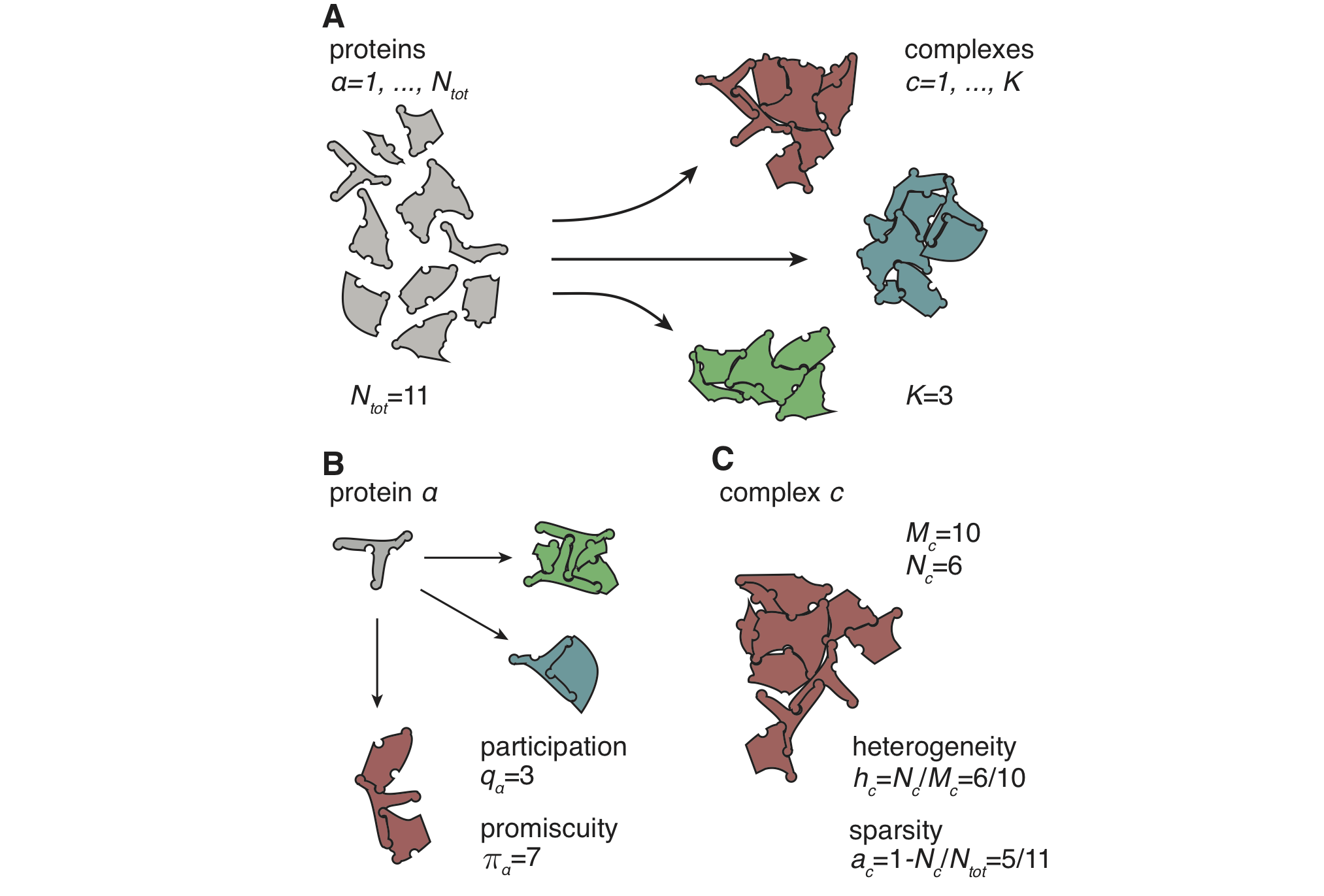}
\caption{{\bf Schematic representation of model parameters.} {\bf A.} Proteins (represented as gray shapes) of $N_{\rm tot}$ different species can assemble into complexes of $K$ different types (each complex type is colored differently). Protein species are labeled $\alpha$, and complex types are labeled $c$. {\bf B.} Each protein species can participate in several complexes, and in each complex can interact with several other species. Accordingly, we define the participation, $q_\alpha$, as the number of complex types in which species $\alpha$ participates; and the promiscuity, $\pi_\alpha$, as the total number of protein species with which species $\alpha$ interacts. {\bf C.} A complex of type $c$ is characterized by its size, $M_c$, which is the total number of proteins it contains, and the number of different protein species it contains, $N_c$. We also define the heterogeneity of the complex, $h_c=N_c/M_c$, as the ratio of composition to size; and its sparsity, $a_c=1-N_c/N_{\rm tot}$, which measures the relative usage that the complex makes of the available protein species.
}
\label{fig:het_reu}
\end{figure}

\begin{figure}
\centering
\includegraphics{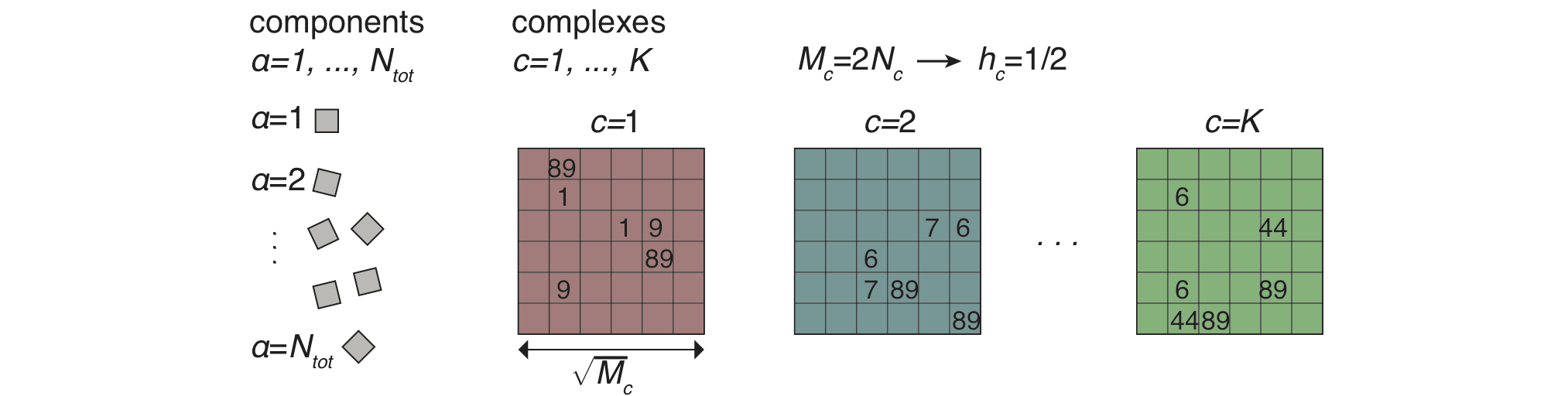}
\caption{{\bf Lattice implementation of the model.} Complexes were generated by randomly sampling the pool of $N_{\rm tot}$ component species and arranging them in squares of side $\sqrt{M_c}$. For $K$ complexes, this procedure was repeated $K$ times. As a  result, some component species participated in several complexes ($\alpha=89$ and $\alpha=6$ in this example). All complexes were constrained to have the same heterogeneity ($h_c=1/2$ in this example).}
\label{fig:gen_comp}
\end{figure}

To support our analytical results, we performed grand canonical Monte Carlo simulations of a lattice implementation of our model, see Fig.~\ref{fig:het_reu} for a summary of notation. We begin by describing how complexes were generated, see also Fig.~\ref{fig:gen_comp}. We considered each of the $c=1,\ldots,K$ different complexes to be a square arrangement of square components representing proteins  with lateral size $\sqrt{M}$, so that $M_c= M$ (all complexes have the same size). Each complex contains $N$ different component species, $N_c= N\le M$ (all complexes have the same number of species). The total number of component species is $N_{\rm tot}\ge N$, and for each complex its $N$ different component species were randomly selected  among the $N_{\rm tot}$. This ensured a sparsity of $a_c= a= 1-N/N_{\rm tot}$, and allowed complexes to share some components with each other. Once the composition of each complex was decided, the spatial arrangements of the components in each complex was determined in the following way. For each complex, each of the corresponding $N$ selected component species was placed in $M/N$ random locations on a square of side $\sqrt{M}$. The completely filled square constituted a complex, see Fig.~\ref{fig:gen_comp}. For the case in which $M/N$ was not integer, the remainder locations on the square were randomly assigned to different component species, such that $N\times\lfloor M/N\rfloor$ species are repeated $\lfloor M/N\rfloor$ times in each complex, and $M-N\times\lfloor M/N\rfloor$ species are repeated $\lfloor M/N\rfloor+1$ times. The result is a heterogeneity $h_c= h\approx N/M$ for all complexes. This procedure allowed us to generate a controlled number of potentially coexisting complexes with given heterogeneity, sparsity, and size.

Following the main text, we defined the interaction energy among component species based on whether two species interact within any of the $K$ different complexes. Note that each complex defines an adjacency bond matrix $B_{\alpha\beta\delta}^c$, where $\alpha$ and $\beta$ are the species that participate in the complex and $\delta=1,\ldots,z$ is the valence, with $z=4$ being the coordination number of the chosen square lattice. The elements of the bond matrix are as follows: if two components of species $\alpha$ and $\beta$ are neighbors with valence $\delta$ in complex $c$, then $B _{\alpha\beta\delta} ^c=1$, otherwise the matrix element is null. We used the bond matrices of complexes to define the interaction energy matrix in the following way:
\begin{align}
U _{\alpha\beta\delta}=-E\Theta\left(\sum_{c=1}^KB _{\alpha\beta\delta} ^c\right)\quad,
\end{align}
where $\Theta$ is the Heaviside function, see also \cite{murugan2015multifarious}. In addition to the binding energy matrix, each species is characterized by its chemical potential, $\mu_\alpha$.

Using $U _{\alpha\beta\delta}$ and $\mu_\alpha$, we could define the energy of the system, $\mathcal{U}$, crucial in Monte Carlo simulations. To do so, we characterized the state of the system by an occupancy matrix $\sigma_{\alpha i}$, which takes the value $1$ if the $i-$th lattice site is occupied by species $\alpha$, and zero otherwise. For notational convenience we used the label $\alpha=0$ to denote empty sites, and set $\mu_0=0$ and $U_{0\beta\delta}=U_{\alpha0\delta}=0$. With these definitions we could write the energy of a lattice configuration $\sigma_{\alpha i}$ as 
\begin{align}\label{eq:Ulatt}
\mathcal{U} = \sum_{\langle i,j\rangle}\sum_{\alpha,\beta}U _{\alpha\beta\delta(i,j)}\sigma_{\alpha i}\sigma_{\beta j}-\sum_i\sum_\alpha\mu_\alpha\sigma_{\alpha i}\quad,
\end{align}
where the first sum spans pairs of near neighbors, and the second spans the whole lattice. Note that, besides for $\alpha=0$, the chemical potential of all species is homogeneous throughout the lattice and among all species, and so $\mu_{\alpha}=\mu$ for $\alpha>0$. Note also that our sign convention is such that $E>0$, i.e. $U_{\alpha\beta\delta}<0$, and $\mu<0$, so that larger and positive $E$ results in stronger binding, whereas lower negative $\mu$ corresponds to lower concentrations (this convention is different from that in \cite{murugan2015multifarious}).

The Monte Carlo algorithm used in the simulations was as follows:
\begin{enumerate}
\item A lattice site $i$ and a species $\alpha$ (including $\alpha=0$) were randomly selected.
\item The change in energy, $\Delta \mathcal{U}$, associated with changing the species currently in $i$ by $\alpha$ was calculated using Eq.~\ref{eq:Ulatt}.
\item The Glauber transition probability was calculated, $W=(1+\exp(\Delta \mathcal{U}))^{-1}$.
\item A random number $p\in [0,1]$ was generated.
\item The species change was performed or not, depending on whether $p<W$ or not.
\end{enumerate}
Steps $1-5$ constituted one Monte Carlo step. Typical simulations were run for $\sim10^6\,{\rm latt}$, where $1\,{\rm latt}$ is a lattice sweep and corresponds to $L$ Monte Carlo steps, with $L=\sqrt{L}\times\sqrt{L}$ the size of the square lattice. Unless otherwise specified, we used $\sqrt{L}=40$. 

In order to quantify our simulations, we used  the density, the energy, and the assembly error as order parameters. The density is defined as the fraction of lattice sites that are occupied (i.e., they contain species $\alpha>0$). The energy of a lattice configuration, $U$ (measured in units of $-E$), is the total number of bonds among near-neighbor lattice sites for which $U_{\alpha\beta\delta}=-E$. Our definition of assembly error is the same as in \cite{murugan2015multifarious}. First, the number of sites, $S$, that match between the target complex and the largest connected cluster in the lattice is computed. Second, the number of sites, $Q$, in the union of the largest connected cluster and the initial seed is also computed. Third, the error is defined as one minus the ratio of these quantities, ${\rm Err}=1-S/Q$.

It is important to note that in the chimeric (Ch) regime this definition of the error does not result in zero error, but rather the error converges to a saturation value ${\rm Err}_{\rm sat}$. This is because in Ch the initial seed remains unchanged and it nucleates chimeric structures that fills the whole lattice. The value of the saturation error thus depends on the initial seed size and on the lateral lattice size $\sqrt{L}$. In particular, if the initial seed has lateral size $\gamma \sqrt{M}$, with $\gamma<1$, then once the whole lattice has been filled by chimeric structures, we have $S=\gamma^2 M$, whereas $Q=L$. Therefore, the saturation error is given by
\begin{align}
{\rm Err}_{\rm sat} = 1 -\gamma^2\frac{M}{L}\quad.
\label{eq:errsat}
\end{align}
In Fig.~2E the initial seed is the whole structure, $\gamma=1$, while the lateral lattice size is $\sqrt{L}=2\sqrt{M}$. This gives ${\rm Err}_{\rm sat} =3/4$, which agrees with the value of the error in Ch, seen in Fig.~2E. In contrast, both in the liquid regime (L) and in the dilute solution regime (DS) the initial seed dissolves, and the error is of order unity, as can also be seen in Fig.~2E. These two latter regimes can however be distinguished by their density, as L is dense and DS is dilute, see Fig.~2F. The multifarious assembly regime (MA), on the other hand,  can be identified by an error value that approaches asymptotically zero, and mean density of $1/4$ for the chosen simulation geometry.

\subsection{Boundaries between regimes in $(E,\mu)$ space}
\label{sec:E_mu_boundary}
The four different regimes of a multifarious mixture that can be observed in Fig.~2 are separated by curves that are roughly linear. We now characterize these curves. For simplicity, we focus on the case $M=N=N_{\rm tot}$ relevant for Fig.~2. In this case the usage of the available component species is ``dense'', and complexes are fully heterogeneous. In the coming sections we will further discuss the roles of heterogeneity and sparsity.

{\bf DS to MA transition.} The low density DS regime is separated from all other regimes by the straight line
\begin{align}\label{eq:mu2}
E=-\mu/2\quad,
\end{align}
where the slope $1/2$ is connected to the geometry chosen in our model, a square lattice. To see this, consider the fully assembled complex as the initial state. The components that are bound more weakly are those at the corners, as each one of them only has two neighbors (unlike three for surface components and four for interior components). These components will be removed (on average) if $2E<-\mu$. Once these are removed, the subsequent surface components will also have two bonds, and they will also be removed. This ``peeling-off'' will continue until the whole complex disassembles, which explains the boundary between DS and L (this is an example of textbook arguments for lattice gas models). 

\begin{figure}
\centering
\includegraphics{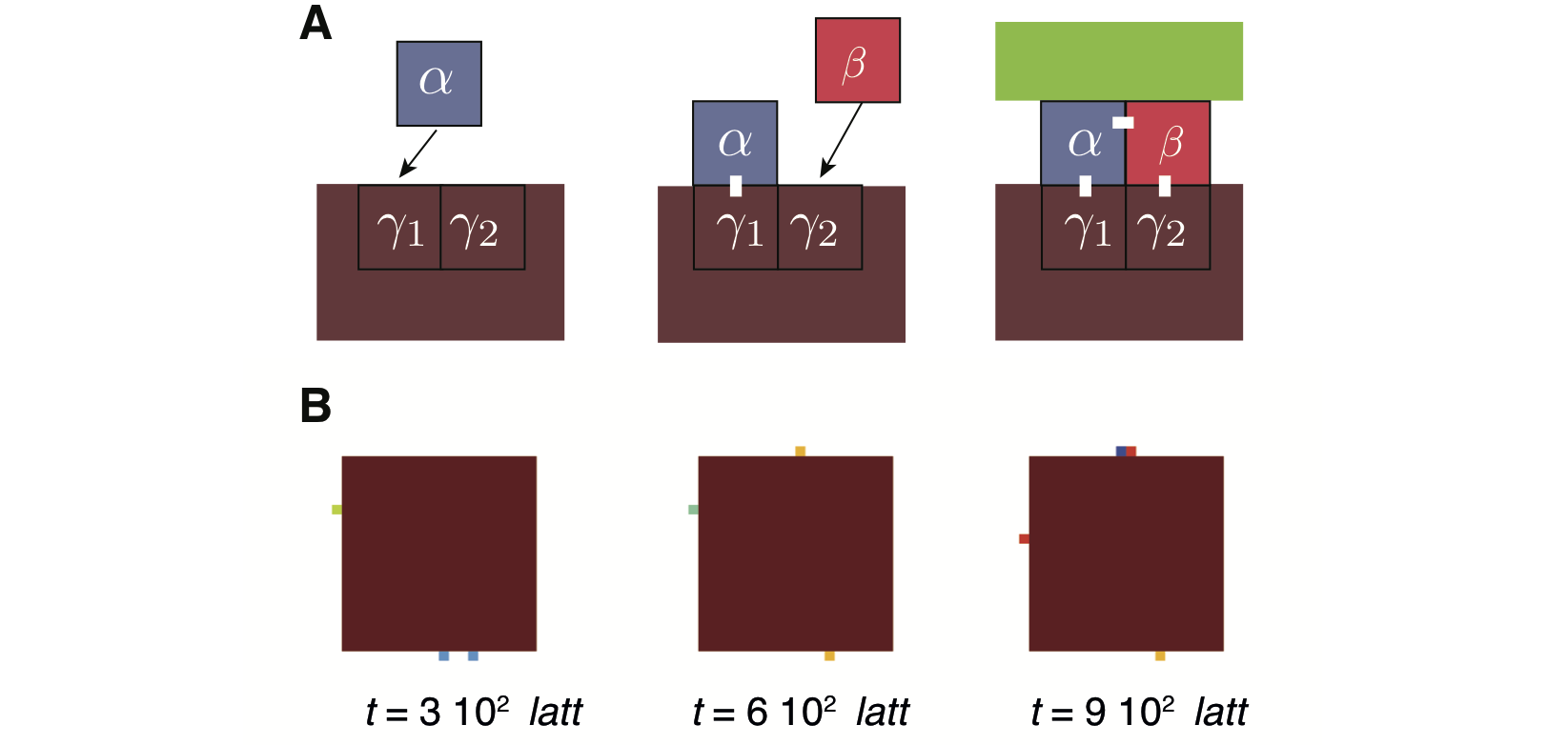}
\caption{{\bf Mechanism for nucleation of chimeric structures.} {\bf A.} A fully assembled complex (colored in brown) contains components $\gamma_1$ and $\gamma_2$ at its surface. A third component, $\alpha$ (blue square), is a component of a different complex, in which it is a neighbor of $\gamma_1$. For this reason, $\alpha$ can establish a bond with $\gamma_1$. Analogously, $\beta$ can form a bond with $\gamma_2$, as they are both partners in a third complex (colored red). In the event that $\alpha$ and $\beta$ are also neighbors in a a fourth complex (colored green), they will also form a bond with each other. Because in this case the number of bonds, three, is larger than the number of components, two, the nucleation seed is stable and will assemble a fragment of the green complex. {\bf B.} Example of the process described in A taking place in an actual simulation with parameters set in the chimeric regime. Three snapshots of the simulation are shown at different times $t$. Notice how in the last snapshot a pair of components belonging to different complexes (colored blue and light red) form a stable seed at the top of the complex (brown square). ({\it Parameters:} $E=12$, $\mu=-17$, $K=8$, $N_c=M_c=N_{\rm tot}=20^2$. Duration $10^6\,{\rm latt}$. Lattice size $40\times40$. Whole complex as initial state.)
}
\label{fig:dyn_chim1}
\end{figure}

{\bf MA to Ch transition.} To explain the onset of Ch, we analyzed simulations, and noted that the event that most likely resulted in chimeric structures was the one depicted in Fig.~\ref{fig:dyn_chim1}A. Consider two neighboring components, of species $\gamma_1$ and $\gamma_2$, at the boundary of an assembled complex, depicted in brown in Fig.~\ref{fig:dyn_chim1}A. Besides participating in the brown complex, these species may also participate in other complexes, in which they will have different neighboring component species. A component, $\alpha$, which does not belong to the brown complex, can thus form a bond with $\gamma_1$ if these two component species, $\alpha$ and $\gamma_1$, are neighbors in a complex different from the brown one, a blue complex in Fig.~\ref{fig:dyn_chim1}A. Similarly, another component $\beta$ can form a bond with $\gamma_2$ if $\beta$ and $\gamma_2$ are neighbors in a third complex, depicted in red in Fig.~\ref{fig:dyn_chim1}A. In principle, $\alpha$ and $\beta$ only have one bond each with the brown complex. However, if they happen to be neighbors in a fourth complex, depicted in green in Fig.~\ref{fig:dyn_chim1}A, then each of them is stabilized by {\it two} bonds. In this case, since $2E>-\mu$, they will form a stable nucleation seed for that fourth complex, which will result in a chimeric structure. In  Fig.~\ref{fig:dyn_chim1}B we show an example of this process observed in a simulation. We can estimate the rate of such event, $W_{\rm chim}$, by multiplying the transition rates for binding $\alpha$ and for binding $\beta$ by the corresponding entropic factors:
\begin{align}\label{eq:chimrate}
W_{\rm chim} \approx \left(4\sqrt{M}\frac{K}{M}\frac{\exp(E+\mu)}{1+\exp(E+\mu)}\right)\left(2\frac{K^2}{M}\frac{\exp(2E+\mu)}{1+\exp(2E+\mu)}\right)\quad.
\end{align}
The first term in parenthesis corresponds to the binding of $\alpha$, and includes two entropic factors and one energetic factor. The first entropic factor $4\sqrt{M}$ arises from the length of the boundary of the existing structure, which we took to be the whole complex. The second corresponds to the chances for component $\alpha$ to be the partner of a given surface component of $c_a$. Because there are $K$ choices for $c_b$ and a total of $M$ different species, this contributes a factor $K/M$. The second term in parenthesis includes two entropic factors: one factor of two, because $\beta$ can be bound to both sides of $\alpha$, and a factor of $K^2/M$, which corresponds to the same factor as before, times $K$ chances that the second bond also matches for the given species $\beta$. The energetic factors correspond to  $\alpha$ establishing only one bond, so that $\Delta\mathcal{U}=E+\mu$, and $\beta$ establishing two bonds, so that $\Delta\mathcal{U}=2E+\mu$.

The transition line between Ch and MA is characterized by high chances of chimeric nucleation, that is $W_{\rm chim}T_{\rm nucl}\approx1$ with $T_{\rm nucl}$ the time needed to nucleate a large chimeric structure. This condition results in a non-linear dependence of $E$ on $\mu$. For large energies (i.e. $E\gg1$, but $E$ and $\mu$ comparable), Eq.~\ref{eq:chimrate} simplifies to $W_{\rm chim} \approx q(K,M)\exp(E+\mu)/(1+\exp(E+\mu))$, where $q(K,M)= 8K^3M^{-3/2}$ is the multiplicity factor. The transition line then becomes
\begin{align}
E=-\mu+\mu_0\quad,
\end{align}
with $\mu_0=-\log(q(K,M)T_{\rm nucl}-1)$. Putting this result together with the transition line between MA and DS, Eq.~\ref{eq:mu2}, we see that the range of chemical potentials for MA is given by $\mu_{\rm max}-\mu_{\rm min}=-E+\mu_0+2E=E+\mu_0$, which is indeed equivalent to Eq.~1 in the main text. To generate the transition line depicted in Fig.~2, we estimated $\mu_0$ by analyzing the Ch-MA transition for $\mu=-30$, and fitting the sigmoidal $g(E)= {\rm Err}_{\rm sat} + (1-{\rm Err}_{\rm sat}) / (1 + \exp(k(E+\mu_0)))$. This resulted in $\mu_0=-10.27\pm0.02$, from which we obtain a value $T_{\rm nucl}\approx6\, \cdotp10^4$.

\begin{figure}
\centering
\includegraphics{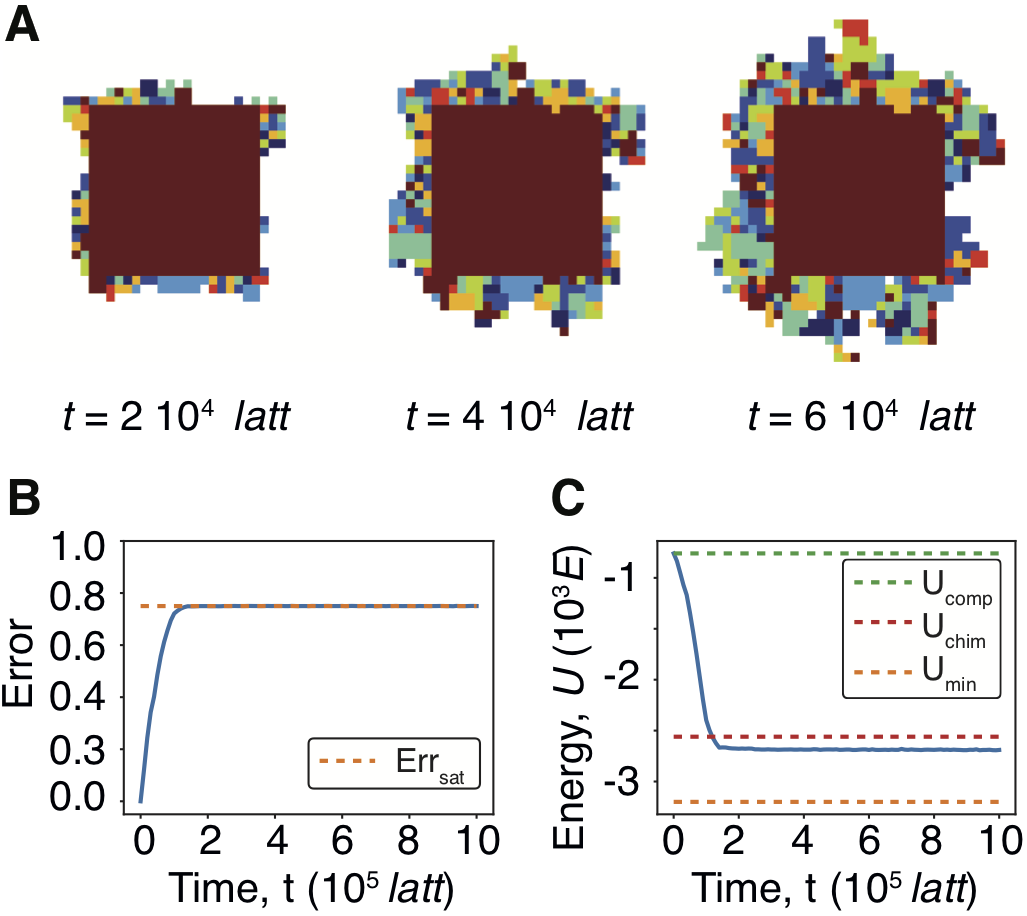}
\caption{{\bf Dynamics of chimeric growth.} 
{\bf A.} Example of chimeric growth in an actual simulation with parameters set in the chimeric regime. Three snapshots of the simulation are shown at different times $t$. Fragments with different colors belong to different complexes.
{\bf B.} Characterization of chimeric growth by the assembly error. The increase in error occurs at $t\sim10^5\,{\rm latt}$, in agreement with the estimate of the nucleation time $T_{\rm nucl}\approx6\,\cdotp10^4$ made in the text.
{\bf C.} Same simulation as in B, characterized by the energy of the system, $U$. Initially the energy coincides with the energy of a complex, $U_{\rm comp}$, but with time it decreases towards the energy of the chimeric regime, $U_{\rm chim}$, always staying above the minimum energy $U_{\rm min}$.
({\it Parameters:} $E=12$, $\mu=-17$, $K=8$, $N_c=M_c=N_{\rm tot}=20^2$. Duration $10^6\,{\rm latt}$. Lattice size $40\times40$. Whole complex as initial state.)
}
\label{fig:dyn_chim2}
\end{figure}

In Fig.~\ref{fig:dyn_chim2} we show an example of the dynamics of assembly in the  chimeric regime. As one can see, the appearance of large assembly errors due to chimeric  growth roughly takes place at the estimated nucleation time $T_{\rm nucl}$. This further supports our analysis. In particular, in Fig.~\ref{fig:dyn_chim2}C we quantify the dynamics of the system energy, $U$. As stated above, $U$ is the number of bonds between components that are specific, i.e. that correspond to the assembling complex. For an isolated complex, the energy is $U_{\rm comp}=-4E(\sqrt{M}\times\sqrt{M}-\sqrt{M})/2$, where $4$ is the number of matching bonds for each component, and $2$ ensures that we are not counting bonds twice. This value matches the initial state in Fig.~\ref{fig:dyn_chim2}C. As the system evolves, more components are added, and the system decreases its (negative) energy as it becomes more stable. The maximum number of specific bonds that can be generated is $U_{\rm min}=-2E\sqrt{L}\times\sqrt{L}$, with $L$ the system size. In Fig.~\ref{fig:dyn_chim2}C we can indeed see that the negative energy of the system stays always below $U_{\rm min}$. Eventually, the system ``freezes'' into a chimeric state and the energy saturates to an intermediate value. We estimate the energy of the chimeric state as the energy of a complex plus the energy of the surrounding material: $U_{\rm chim}\approx U_{\rm comp} -E (\sqrt{L}\times\sqrt{L}-\sqrt{M}\times\sqrt{M})f/2$. Here, the factor $f$ corresponds to the average number of bonds in the material surrounding the complex. For the case in which all bonds match corresponding to MA, $f=4$. At the other extreme, when growth occurs through filamentous structures, we have $f=2$. In the case presented in the figures we used the intermediate value $f=3$.

\begin{figure}
\centering
\includegraphics{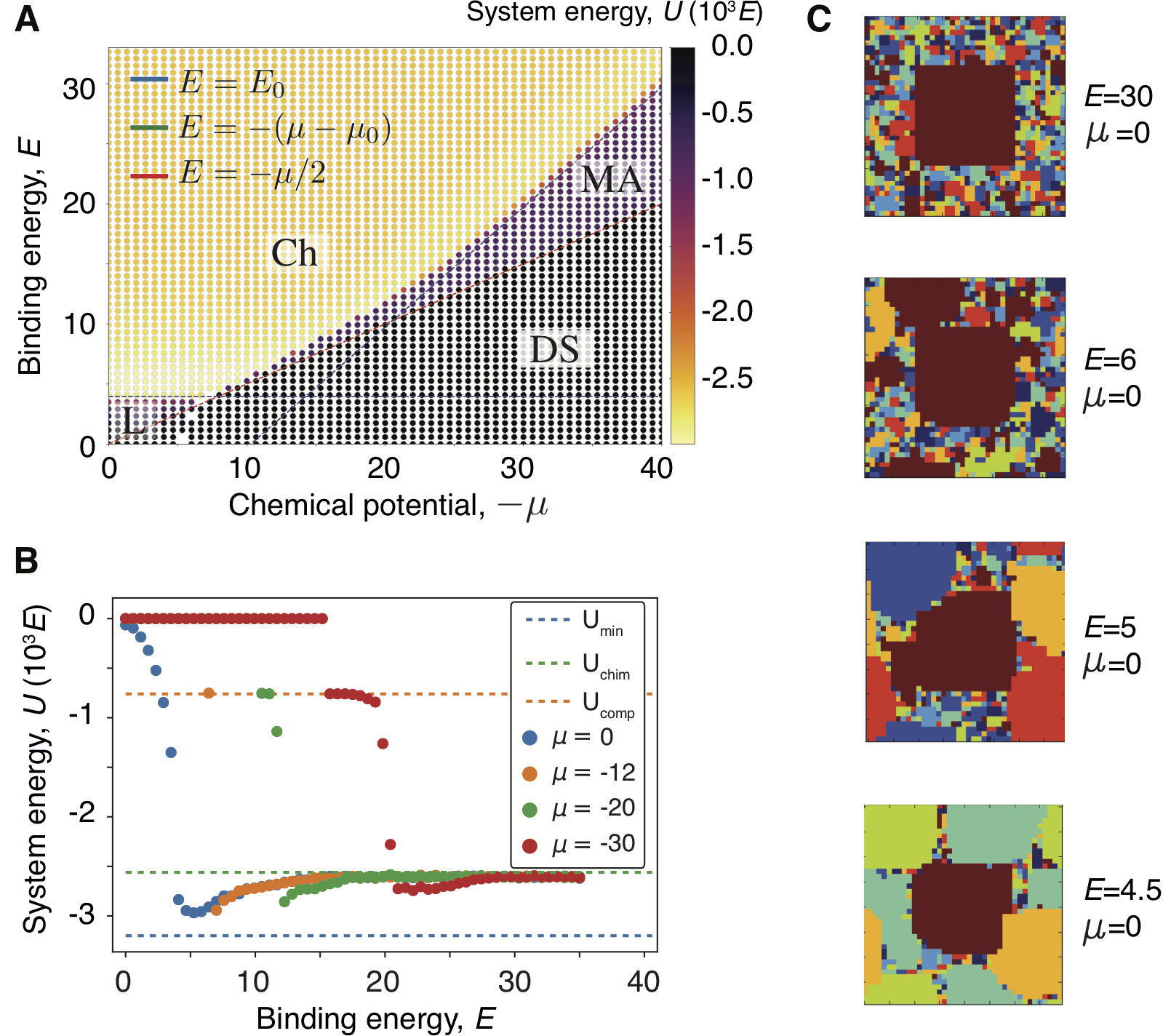}
\caption{{\bf Characterization of the Ch-L transition.} 
{\bf A.} Regime diagram using the energy of the system, $U$, as order parameter. Regime boundaries are the same as in Fig.~2. The L and DS regimes are characterized by high system energies, the Ch regime by low system energy, and the MA regime by intermediate system energy. Note that the transition into Ch is gradual, with lower energies near the boundaries (on the Ch side) than deep inside the Ch regime. 
{\bf B.} The system energy is non-monotonic in the transition from liquid to chimeric regimes. This is because near the boundary the binding energy, $E$, is low enough so to allow rearrangements of the disordered components into macroscopic fragments of complexes. This allows the system to reach a lower energy state.
{\bf C.} Examples of structures in Ch far (above) and near (below) the transition into L. As $E$ is lowered towards its transition value (on the Ch side), large fragments of complexes assemble. For large $E$ the time-scale for the formation of these fragments is much larger than the simulation time.
({\it Parameters:} $K=8$, $N_c=M_c=N_{\rm tot}=20^2$. Duration $10^6\,{\rm latt}$, lattice size $40\times40$, whole complex as initial state. Panels in C correspond to last simulation step.)
}
\label{fig:glass_liq}
\end{figure}

{\bf Ch to L transition.} This transition occurs at low values of $E$ and shows no dependence on $\mu$, see Fig.~2E and F. To understand why this is the case, note that because both regimes are dense, components are replaced by other components without any role for empty sites, therefore the transition rate is independent of $\mu$. We determined the transition energy $E_0$ by fitting on the $\mu=0$ line the sigmoidal function $g(E)= {\rm Err}_{\rm sat} / (1 + \exp(k(E-E_0)))$. This resulted in $E_0=4.0\pm0.1$. Although we do not extensively characterize the boundary between Ch and L, in Fig.~\ref{fig:glass_liq}A we show how the onset of L is preceded by a decrease in the energy of the system. This is further quantified in Fig.~\ref{fig:glass_liq}B, e.g. for the $\mu=0$ curve. The reason for the observed non-monotonicity is that at low binding energies the state of the system is not fully frozen, and components can rearrange over time to form large fragments of complexes separated by disordered boundaries. Examples of such behavior for different values of the energy through the transition are depicted in Fig.~\ref{fig:glass_liq}C.

\subsection{Scaling laws}
\label{si:scale}
In the previous section we discussed the boundaries of the different regimes of our model in the thermodynamic space $(E,\mu)$. The position of these boundaries  depend on the parameters that characterize the complexes, $(h,a,K,M)$, through algebraic relations. Such dependencies result in the scaling laws of the main text, which we now derive.

{\bf Estimating promiscuity.} We begin by providing an estimate for the promiscuity of components. First, we note that a given component can bind $z$ other components simultaneously, and it thus has $z$ different binding sites. We now study the promiscuity per binding site of a given species $\alpha$. Because on average the promiscuity of all components is the same, hereafter we note $\pi_\alpha=\pi$. We first discuss the case of a mixture that only consists of components from a single complex ($K=1$ and $N=N_{\rm tot}$). Importantly, the dependence of the promiscuity on the heterogeneity is non-monotonic. For very low heterogeneities, $N\ll M$, the promiscuity is roughly equal to the number of species, $\pi\approx N=hM$, i.e. all species interact with all species. This is because the probability that none of the $M/N$ members  of one species interacts with another given species in the structure is $P_{\rm no\;int}\approx(1-1/N)^{M/N}$, which for $N\ll M$ is neglible. However, as $N$ increases and becomes comparable to $M$, the chances of two species to not interact becomes of order one. We can estimate the value of the heterogeneity at the cross-over between these two cases, $h_0$. This is done by expanding  asymptotically in the system size, $\epsilon\equiv M^{-1}$,
\begin{align}
P_{\rm no\;int}\approx(1-1/N)^{M/N}=(1-\epsilon/h)^{1/h}\approx1-\epsilon/h^2+\mathcal{O}(\epsilon^2)\quad.
\end{align}
We see now that for
\begin{align}\label{eq:h0}
h\gtrsim h_0\approx M^{-1/2}
\end{align}
the second term becomes small. In this case, the probability that there are no interactions becomes of order one, and our previous estimate breaks down.

\begin{figure}
\centering
\includegraphics{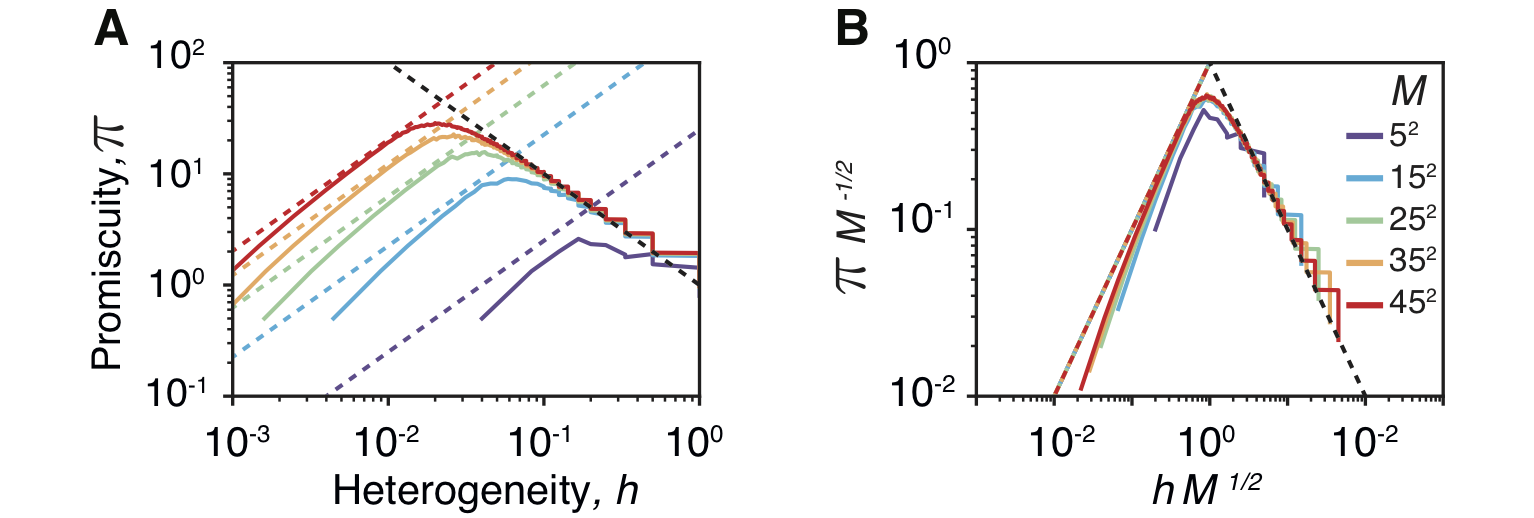}
\caption{{\bf Dependence of promiscuity on heterogeneity.} {\bf A.} Solid lines represent the average promiscuity of components in computationally generated complexes with different heterogeneities. Each color corresponds to complexes of a different size (see legend in B). Colored dashed lines represent the low-heterogeneity estimate, $\pi\approx hM$, and black dashed line the high heterogeneity estimate, $\pi\approx h^{-1}$. These analytical estimates agree well with the computational results.
{\bf B.} Same as A, but after rescaling the heterogeneity by $h_0=M^{-1/2}$ and the promiscuity by $M^{1/2}$. The collapse of the computational results agrees with the analytical estimates. }
\label{fig:combinatorics}
\end{figure}

We can obtain a different estimate of the promiscuity for large values of the heterogeneity. If the complexes are fully heterogeneous, then each species will interact just with another one, and so the promiscuity is $\pi=1$. Reducing the heterogeneity increases the number of members of each species present in the complex. If each of these species interacts with a different species, we have $\pi\approx M/N=h^{-1}$. It is this second estimate of promiscuity that we used in the main text. The justification for the choice of this second estimate is as follows. The scaling of the cross-over heterogeneity is $h_0\sim M^{-1/2}$, while the scaling of the heterogeneity value that determines the MA to Ch transition is $h\sim M^{-1/3}$ (see Eq.~\ref{eq:si_trans1} below). Therefore, for the values of $h$ in which the second estimate of the promiscuity breaks down and the first estimate should be used instead, the system is already in the chimeric regime. In Fig.~\ref{fig:combinatorics}A and B we show how randomly reshuffling $M$ components of $N$ different types indeed results in an average promiscuity that behaves as described.

We now generalize the second estimate of promiscuity for the case in which there are several complexes and the usage of the component species is sparse, i.e. $K>1$ and $a<1$. In this case, for randomly sampled complexes, the promiscuity will be decreased by the probability that a given component species is selected, $N/N_{\rm tot}=1-a$, and increased by the number of total complexes, $K$. We thus have
\begin{align}\label{eq:prom}
\pi\approx K (1-a)/h\quad,
\end{align}
which is used in the main text. We stress that this estimate holds for randomly generated complexes of size $M$ with heterogeneity larger than $h_0\approx M^{-1/2}$.

{\bf MA to Ch scaling.} We can use the estimate of the promiscuity in Eq.~\ref{eq:prom} to obtain an estimate of the maximum number of coexisting complexes that can be reliably assembled. To do so, consider a square seed of a complex on a square lattice, i.e. $z=4$ (we will later generalize to arbitrary $z$). The components added to the edge of a growing layer (for a square complex, the edge is a line of components) have two near neighbors. Each of the two components in the growing tip of the complex defines a set of $\pi$ species, with which the component to be added can have specific interactions. For small $\pi$, the probability that these two sets intersect, $P_{\cap}$, is small: the only intersection corresponds to the correct species for the given complex. Therefore, for low promiscuity assembly errors are scarce. On the other hand, as $\pi$ increases the probability of these two sets to intersect can become of order unity. The seed starts then to  grow into a chimeric structure. We shall now estimate $P_{\cap}$.

We have $P_{\cap} = 1-P_{\neg\cap}$, and for random structures, $P_{\neg\cap}$ can be estimated as the probability that each of the $\pi$ components $\alpha$ included in the first set are not in the second, and so $P_{\neg\cap}=\prod_\alpha P_{\neg\cap_\alpha}=P_{\neg\cap_\alpha}^{\pi}$. Estimating the probability that a component $\alpha$ is in the second set as $P_{\cap_\alpha}\approx{\pi}/{N_{
\rm tot}}$, we arrive at
\begin{align}
P_{\cap} \approx 1-\left(1-\frac{\pi}{N_{
\rm tot}}\right)^{\pi} =1-\left(1-\epsilon\frac{\pi(1-a)}{h}\right)^{\pi} = \epsilon \frac{\pi^2(1-a)}{h}+\mathcal{O}(\epsilon^2)=\epsilon \frac{K^2(1-a)^3}{h^3} +\mathcal{O}(\epsilon^2) \quad,
\end{align}
where in the second equality we have used the definitions of heterogeneity and sparsity, and --as before-- we have taken $\epsilon=M^{-1}$. To calculate the scaling for the boundary separating the Ch and MA regimes, we set $P_{\cap}\lesssim1$. This results in
\begin{align}\label{eq:si_trans1}
K^2\lesssim \frac{h^{3}M}{(1-a)^3}\quad.
\end{align}
Setting in this equation $a=0$ and $K=1$, we recover $h\gtrsim M^{-1/3}$, while setting only $a=0$ we obtain $K\lesssim h N_{\rm tot}^{1/2}$. Generalization to an arbitrary coordination number $z$ requires to note that the surface components have in general $z/2$ neighbors. Our previous estimate is then modified by considering the intersection of $z/2$ sets, and so
\begin{align}
P_{\cap} \approx 1-\left(1-\frac{\pi^{z/2-1}}{N_{\rm tot}^{z/2-1}}\right)^{\pi}= \epsilon^{z/2-1}\frac{\pi^{z/2}(1-a)^{z/2-1}}{h^{z/2-1}}+\mathcal{O}(\epsilon^{z-2})=\epsilon^{z/2-1}\frac{K^{z/2}(1-a)^{z-1}}{h^{z-1}}+\mathcal{O}(\epsilon^{z-2})\quad,
\end{align}
which through an expansion in $\epsilon$ directly results in
\begin{align}
K^{z/2}\lesssim\frac{h^{z-1}}{(1-a)^{z-1}}M^{z/2-1}\quad.
\end{align}

We can reconcile these scaling relations with the entropic factors found for the rate for chimeric structure formation, Eq.~\ref{eq:chimrate}, in the following way. Consider that a point in $(E,\mu)$ within the MA regime is chosen. If we start increasing the number of complexes, the entropic factor $q(K,M)$ of $W_{\rm chim}$ in Eq.~\ref{eq:chimrate} will increase algebraically, shifting the offset $E_0$ of the boundary with the Ch regime, see Fig.~2. Eventually, the point chosen in $(E,\mu)$ will fall in Ch. Note that Eq.~\ref{eq:chimrate} predicts a particular scaling for this shift in the boundary offset, which is precisely $K\sim \sqrt{M}$, analogous to Eq.~\ref{eq:si_trans1} in the limit of $N=M=N_{\rm tot}$.

Throughout the main text we used $z=4$. The reason is that for this value we could carry out numerical simulations on a square lattice that validated our analytical results. We now discuss the validity of our arguments for other values of $z$.
\begin{itemize}
\item First, we note that the heterogeneity scaling, which was $h\gtrsim M^{-1/3}$ for $z=4$, is in general given by $h\gtrsim M^{(2-z)/(2z-2)}$. The size scaling exponent is therefore negative for $z>1$ (it diverges for $z=1$), and is always larger than $-1/2$. Because $h_0$ has a size scaling exponent of $-1/2$, see Eq.~\ref{eq:h0}, the latter ensures that the estimate for the promiscuity that was used, Eq.~\ref{eq:prom}, is adequate all throughout the MA regime. The fact that, for $z>1$, the lower bound for $h$ has a negative size scaling exponent  allows us to conclude that in all cases there is a lower bound for heterogeneity that decreases with system size. Therefore, our argument about the role of high heterogeneity in assembly should remain unaffected for $z>1$. Note that the only complexes for which $z=1$ are dimeric complexes, which we do not expect to obey our scaling predictions due to their small size.

\item The combinatorial expansion for a dense usage of the proteome, $a=0$, is in general bounded by $K\gtrsim h N_{\rm tot}^{(z-2)/z}$. This results in a sub-linear combinatorial expansion for $z>2$. Therefore, the argument of weak combinatorial usage is valid for $z>2$. The case $z=2$, which corresponds to filamentous complexes, in which each component has two neighbors, does not allow for any combinatorial expansion, as already noted in \cite{murugan2015multifarious}.

\item Finally, allowing for a sparse usage of components, we obtain $K\gtrsim N_{\rm tot}^{(2z-2)/z}M^{-1}$, and the super-linear scaling that results from sparsity is preserved for all values $z>2$. Again, the case $z=2$ is special, as it results in a linear scaling with the proteome size.\end{itemize}

These arguments show that the three main conclusions of our work should remain valid for $z>2$: ($i$) higher heterogeneity reduces chances of chimeric structure formation, ($ii$) combinatorial expansion is weak for dense proteome usage, ($iii$) sparsity allows for super-linear combinatorial expansion. The case $z=2$, which corresponds to filaments, has to be discussed separately. Within our framework, for $z=2$ only fully heterogeneous complexes (filaments) can be assembled. To see this, consider a filamentous complex, in which one species $\alpha$ appears twice, while the rest appear only once. If $\alpha$ is bound to the growing tip of the complex, there are two possible partners that can bind to it with equal strength, which would result in incorrect assemblies with probability of one half. For the same reason, components can not be shared. Therefore, for $z=2$ a dense usage of the proteome only allows for a single complex. A sparse usage clearly allows for a number of complexes that, in order not to share components, must increase linearly with the proteome size. Therefore, in this case sparsity is also (rather trivially) necessary. 

\begin{figure}
\centering
\includegraphics{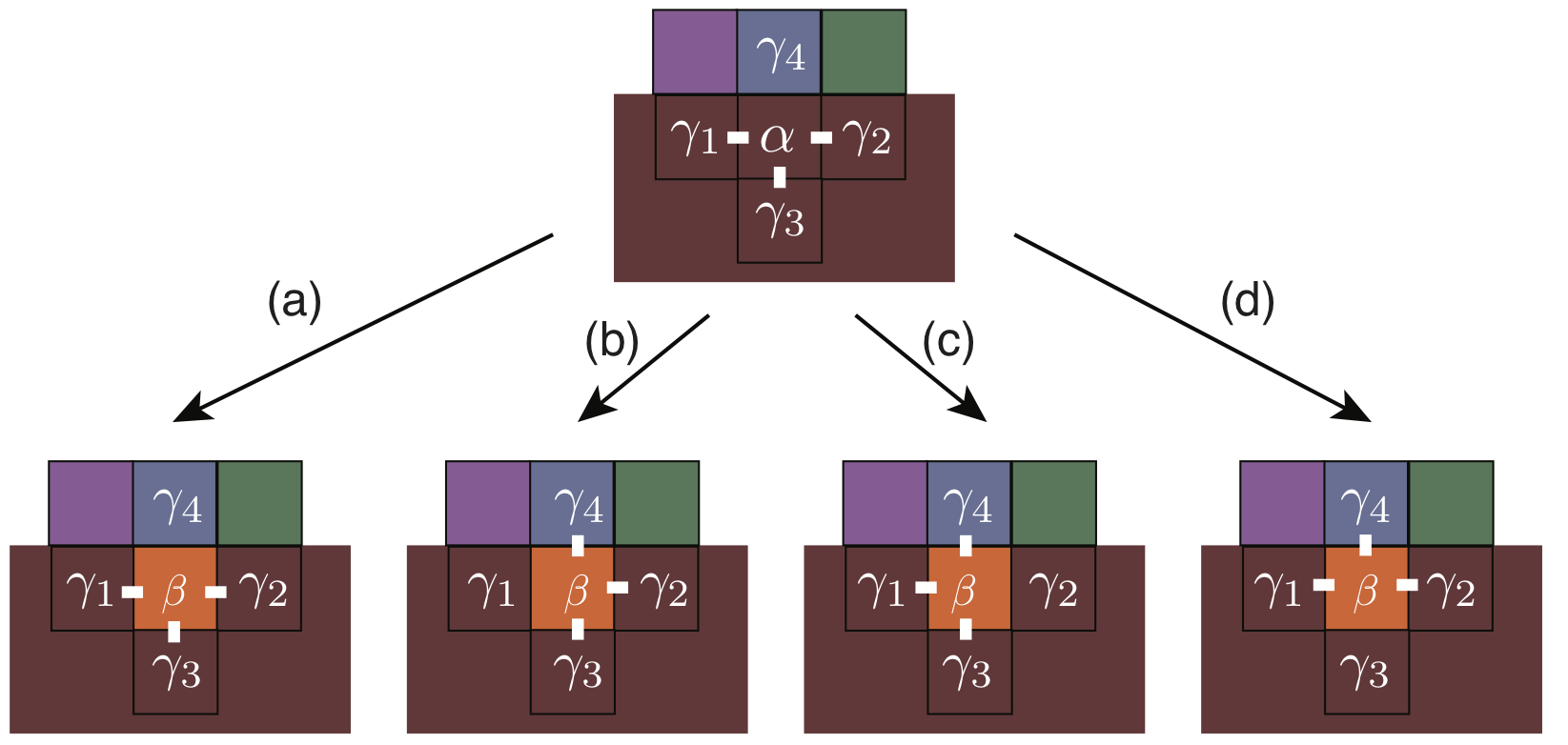}
\caption{{\bf Schematic of the mechanism leading to the dissolution of a complex surrounded by a chimeric structure.} A complex, depicted in brown, is surrounded by components that form a chimeric structure (purple, blue and green components belong to fragments of different complexes). A component at the surface of the complex, such as $\alpha$, is typically stabilized by three bonds (bonds are represented with white bold lines). If the promiscuity of components is sufficiently large, $\alpha$ can be replaced by another component, $\beta$, which also forms three stable bonds. If the bonds of $\beta$ are the same as those of $\alpha$, as is the case in (a), the complex will remain stable. However, if the bonds are established with different components, as shown in (b), (c) and (d), the replacement of $\alpha$ by $\beta$ will reduce the number of bonds of other components in the complex ($\gamma_1$, $\gamma_2$ or $\gamma_3$, respectively), and hence destabilize the complex. This will eventually lead to the dissolution of the complex.
}
\label{fig:dyn_liq}
\end{figure}

{\bf Ch to L scaling.} The scaling of the transition from Ch to L can be estimated with a similar argument as that of the transition from MA to Ch. We consider a full complex surrounded by a disordered chimeric structure, and our goal is to determine under what conditions the disordered aggregate will make the complex vanishg. At the interface with the complex, only some of the components in the aggregate (those that nucleated the chimeric structures, such as $\alpha$ and $\beta$ in Fig.~\ref{fig:dyn_liq}) have specific interactions with the complex. The majority of the interface components have specific interactions with the aggregate, but not with the complex. Therefore, the majority of the components at the boundary of the complex, such as $\alpha$ in Fig.~\ref{fig:dyn_liq}, are stabilized by three bonds (two with other surface components of the complex, $\gamma_1$ and $\gamma_2$, and one with an interior component of the complex, $\gamma_3$), and have frustrated interactions with the interface components, $\gamma_4$. If the promiscuity is very large, then a component $\beta$, non-specific to the complex, may satisfy the compatibility condition established by three of the four neighboring components, and thus replace $\alpha$ at no energetic cost. If the three neighbors to which $\beta$ binds are the same as $\alpha$, the process stops there. But if one of the neighbors corresponds to the disordered aggregate, $\gamma_4$, then one of the components of the complex will loose the bond, see Fig.~\ref{fig:dyn_liq}. This component will then become be less stable, and more susceptible of being replaced by an alternative component that does not belong to the complex. This process will eventually result in the disordered aggregate ``dissolving'' the complex, as seen in Fig.~2C. To estimate the scaling that corresponds to this process we note that $\beta$ establishes three specific bonds, and so it is at the intersection of three sets of size $\pi$. The probability that these sets intersect, $P_\cap$, is now given by
\begin{align}
P_{\cap} \approx 1-\left(1-\frac{\pi^{2}}{N_{\rm tot}^2}\right)^{\pi}= \epsilon^2  \frac{K^3(1-a)^5}{h^5}+ \mathcal{O}(\epsilon^{4})\quad,
\end{align}
which by setting $P_{\cap}\approx1$ directly gives
\begin{align}\label{eq:si_trans2}
K^3\lesssim \frac{h^{5}M^{2}}{(1-a)^5}
\end{align}
for the case $z=4$. Generalization of this expression to arbitrary  $z$ requires noting that an an outer surface of a $z/2$ dimensional complex has $z-1$ neighbors conditioning each of its components. Therefore, it is the intersection of $z-1$ sets that has to be considered, which gives $K^{z-1}\lesssim h^{2z-3}M^{z-2}/(1-a)^{2z-3}$.

\subsection{Finite size scaling}
\label{sec:exponents}

The transition from MA into Ch is characterized by Eq.~\ref{eq:si_trans1}, and that from Ch into L by Eq.~\ref{eq:si_trans2}. These equations predict particular size scaling relations of $a$, $h$, and $K$ with $M$. Denoting with asterisk subscripts the values at the boundaries, we define the following exponents:
\begin{align}
{\rm MA}\to {\rm Ch}:\quad&K_*\sim M^{\nu}\quad;\quad h_*\sim M^{-\mu}\quad;\quad 1-a_*\sim M^{\gamma} \\
{\rm Ch}\to {\rm L}:\quad&K_*\sim M^{\bar{\nu}}\quad;\quad h_*\sim M^{-\bar{\mu}}\quad;\quad 1-a_*\sim M^{\bar{\gamma}}
\end{align}
With these definitions, Eq.~\ref{eq:si_trans1} implies $\nu=1/2$, $\mu=1/3$, $\gamma=1/3$, while Eq.~\ref{eq:si_trans2} implies $\bar{\nu}=2/3$, $\bar{\mu}=2/5$, $\bar{\gamma}=2/5$.  To numerically confirm these analytical predictions, we performed three classes of numerical simulations, which we describe below. In order not to make any assumption about the form of the scaling function, we used the procedure developed in \cite{houdayer2004low}. In short, consider $s_{\rm tot}$ datasets $D_1,\ldots,D_{s_{\rm tot}}$, each corresponding to a different system size (here complex size, $M$) and each composed by a set of points $i=1,\ldots,t_{s}$. For each of the points there exists a value of the scaling parameter, $x_i$ (for example, the heterogeneity $h$); a value of the function, $g_i$, whose scaling we want to study (e.g., the assembly error); and a standard deviation for the value of the function, $\delta_i$ (here calculated from different replicate simulations). That is, we have $s_{\rm tot}$ datasets $D_s = \{x_i,g_i,\delta_i\}_1^{t_s}$, which in our case are generated through numerical simulations. Provided these datasets, the procedure in \cite{houdayer2004low} provides a quality function, $S(\mu)$,  of the scaling exponent, $\mu$. This quality function measures the mean square distance of the rescaled values of the function in units of standard errors. The estimated value of the exponent is obtained by minimizing the quality function, with a good data rescaling corresponding to values below $S=2$. The error in the exponent is obtained by calculating the value of the deviation in exponent for which the quality function increases by one.

Three classes of simulations were performed. In all of them we considered a $L=40\times40$ lattice, and studied assembly sizes, $M=10^2,11^2,\ldots,32^2$ with the binding energy set to $E=7$ and the chemical potential to $\mu=12.6$. Sixty replicate simulations were done for each parameter set, each with a duration of $10^6{\rm latt}$. To determine each exponent, we used as the order parameter the error at the end of the simulation. We also used the saturation error in the Ch regime, given by Eq.~\ref{eq:errsat}, to rescale the error as explained below (only rescaled errors below $1.05$ were considered). We would like to emphasize that the range of assembly sizes that we considered in the simulations is {\it insufficient} to accurately determine the scaling of the regime boundaries. To accurately determine these scalings would require expanding the range of sizes to cover at least two orders of magnitude. The purpose of our simulations is  simply to provide numerical support for our analytical arguments, i.e. for Eqs.~\ref{eq:si_trans1} and \ref{eq:si_trans2}. Importantly, cellular protein complexes have sizes of at most $M\sim 10^2$, so the precise value of the exponents is unlikely to be biologically relevant.

In the first class of simulations we fixed the heterogeneity to $h=0.9$ and the sparsity to $a=0$, i.e. $N=N_{\rm tot}$. We then changed the number of coexisting complexes, $K$, for various sizes of complexes, $M$, in order to determine $\nu$ and $\bar{\nu}$. It is convenient to introduce the loading variable $n\equiv K/M$.

\begin{figure}
\centering
\includegraphics{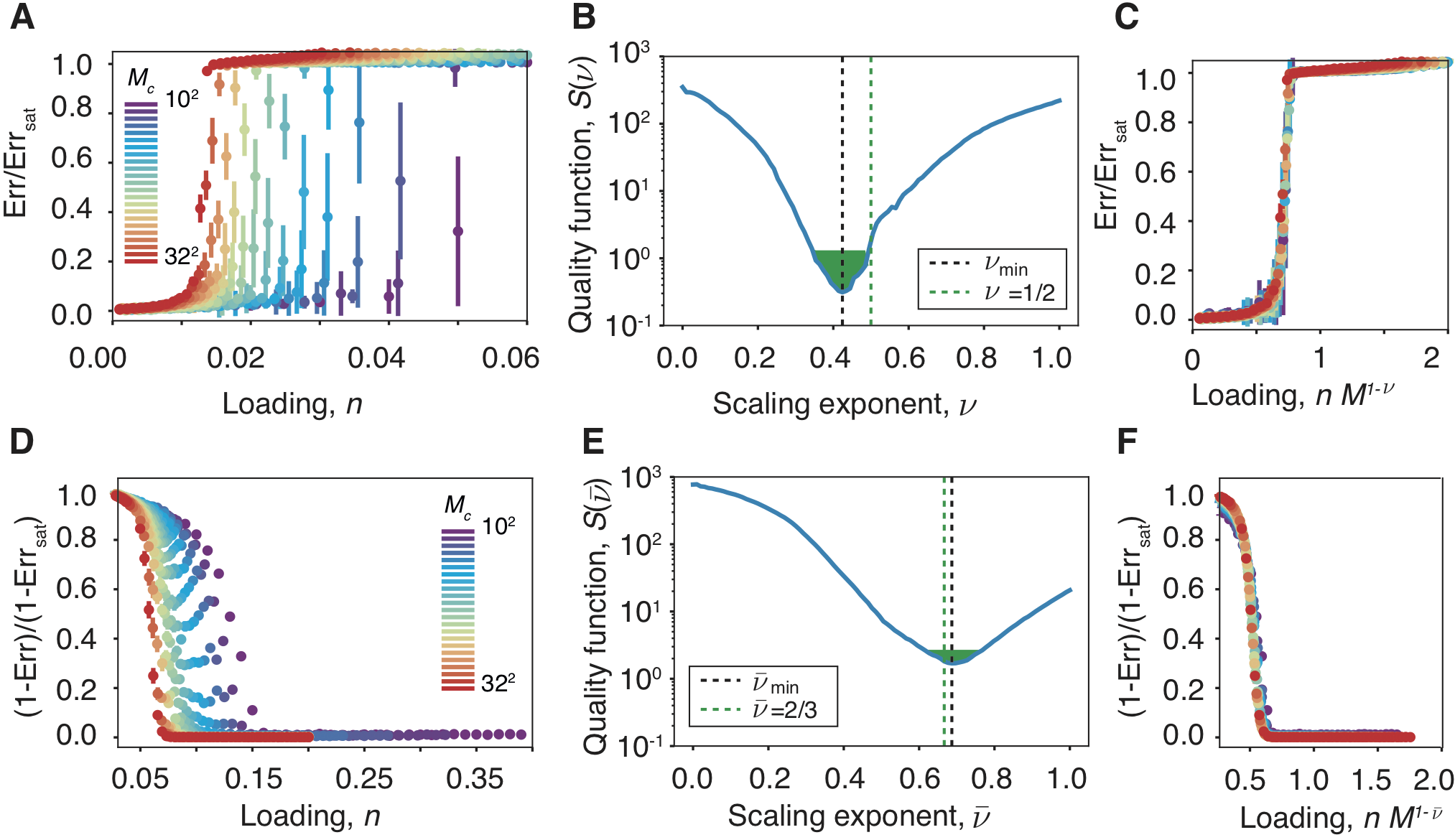}
\caption{
{\bf Scaling analysis of the loading, $n\equiv K/M$.}
{\bf A.} Assembly error as a function of the loading. Each point is an average of $60$ simulations, with bars denoting standard deviations. At high loading there is a transition from MA into Ch, and the error saturates to ${\rm Err}_{\rm sat}$ (see Eq.~\ref{eq:errsat}). Different colors correspond to different sizes, see legend.
{\bf B.} The data is rescaled with different size scaling exponents, ${\nu}$. For each exponent the quality of the data collapse is quantified by the function $S(\nu)$ as defined in \cite{houdayer2004low}, see also text. Optimal data collapse occurs at the minimum, $\nu_{\rm min}=0.42$. We estimate $\delta\nu=0.07$ as the width of the region in which $S(\nu)$ increases by unity, shaded in green. This is in agreement with the analytical estimate $\nu=1/2$.
{\bf C.} Collapse of the data from A using $\nu_{\rm min}$. {\bf D}, {\bf E} and {\bf F.} Analogous to panels A to C, but for the Ch to L transition, with $\bar{\nu}_{\rm min}=0.69$ and  $\delta\bar{\nu}=0.07$, compatible with the analytical estimate $\bar{\nu}=2/3$.
({\it Parameters}:  $E=7$, $\mu=-12.6$, $h=0.9$, $a=0$. Duration $10^6\,{\rm latt}$, lattice size $40\times40$, whole complex as initial state.)
\label{fig:scaling_nu}}
\end{figure}

\begin{itemize}
\item {\bf Determining $\nu$.}  To calculate $\nu$ we used the scaling hypothesis ${\rm Err}/{\rm Err}_{\rm sat}= f(K/K_*)= f(n/n_*)=g(n M^{1-\nu})$, and studied low values of the loading, $n\in[0, 0.06]$, so that the system would only cross the first regime boundary, MA to Ch. This is the transition studied in \cite{murugan2015multifarious}. In Fig.~\ref{fig:scaling_nu}A we see that the error changes smoothly from near $0$ to near its saturation value. Fig.~\ref{fig:scaling_nu}B shows the fit quality function, which takes its minimum at $\nu_{\rm min}=0.42$. By calculating the value for which the quality function increases by unity, we estimate $\delta\nu=0.07$. This matches well the analytical prediction $\nu=1/2$. In Fig.~\ref{fig:scaling_nu}C we can see that with the numerically determined exponent there is a good data collapse.

\item {\bf Determining $\bar{\nu}$.}  To calculate $\bar{\nu}$ we analyzed the same class of simulations as above, but now used the scaling hypothesis $(1-{\rm Err})/(1-{\rm Err}_{\rm sat})= f(K/K_*)= f(n/n_*)=g(n M^{1-\bar{\nu}})$. We studied high values of the loading, $n\in[0.03, 0.4]$, so that the system would cross the second regime boundary, Ch to L. In Fig.~\ref{fig:scaling_nu}D we see that the error changes smoothly from near its saturation value, at the end of the first transition, to near unity, in the liquid regime. Fig.~\ref{fig:scaling_nu}E shows the quality function of the fits, which has a minimum at $\bar{\nu}_{\rm min}=0.69$, with $\delta\bar{\nu}=0.07$. This matches well the analytical prediction $\bar{\nu}=2/3$. In Fig.~\ref{fig:scaling_nu}F we can see that there is a good data collapse.
\end{itemize}

In the second class of simulations we fixed the number of complexes to $K=5$ and the sparsity to $a=0$, i.e. $N=N_{\rm tot}$. We then changed the heterogeneity, $h$, for several complex sizes, $M$, in order to determine $\mu$ and $\bar{\mu}$.

\begin{figure}
\centering
\includegraphics{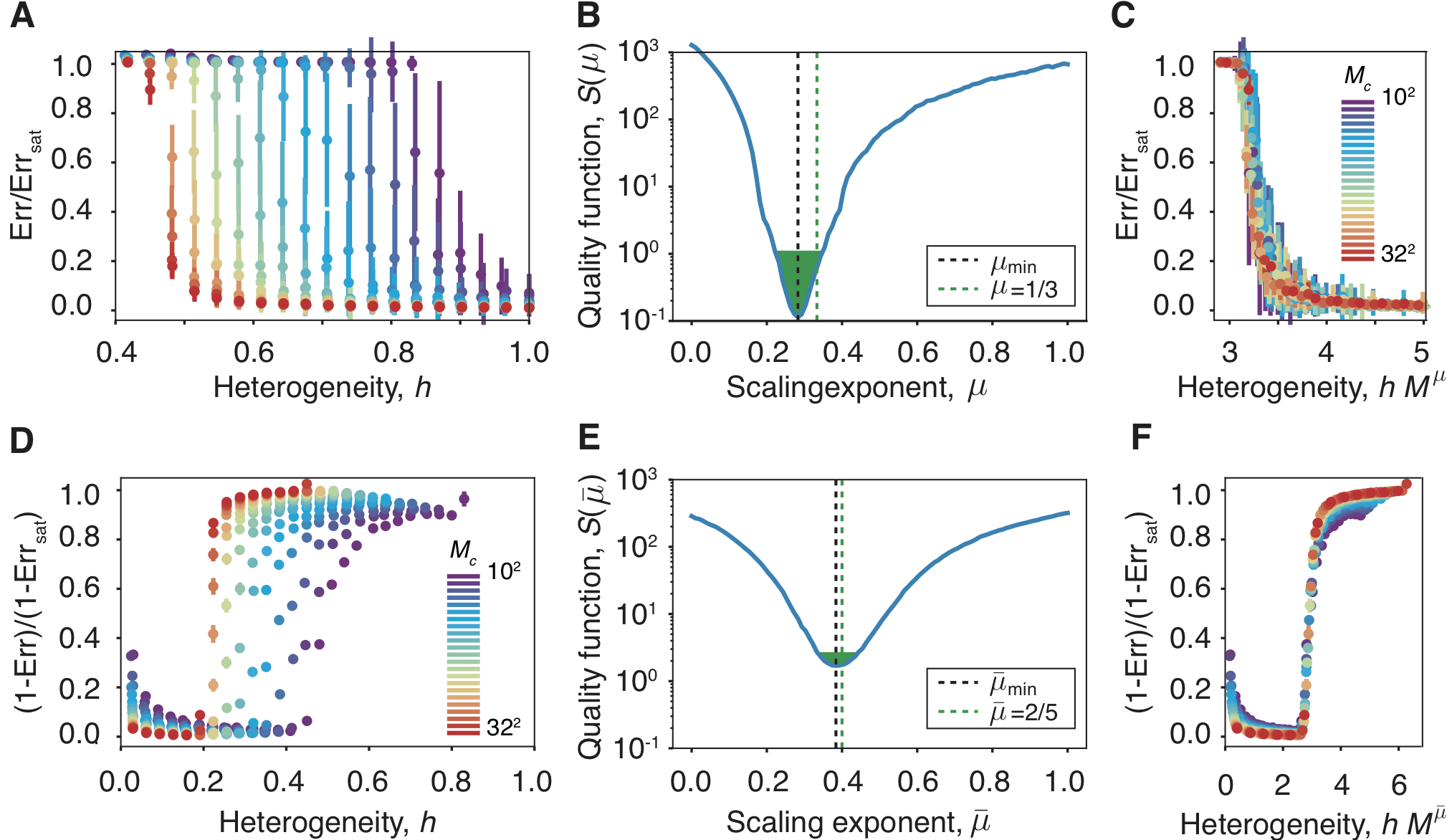}
\caption{
{\bf Scaling analysis of the heterogeneity, $h$.}
{\bf A.} Assembly error as a function of the heterogeneity. Each point is an average of $60$ simulations, with bars denoting standard deviations. At low heterogeneity there is a transition from MA into Ch, and the error saturates to ${\rm Err}_{\rm sat}$ (see Eq.~\ref{eq:errsat}). Different colors correspond to different sizes, see legend in panel C.
{\bf B.} The data is rescaled with different size scaling exponents, ${\mu}$. For each exponent the quality of the collapse is quantified by the function $S(\mu)$ as defined in \cite{houdayer2004low}, see also text. Optimal data collapse occurs at the minimum, $\mu_{\rm min}=0.28$. We estimate $\delta\mu=0.06$ as the width of the region in which $S(\mu)$ increases by unity, shaded in green. This is in agreement with the analytical estimate $\mu=1/3$.
{\bf C.} Collapse of the data from A using $\mu_{\rm min}$. {\bf D}, {\bf E} and {\bf F.}  Same as A to C, but for the Ch to L transition, with $\bar{\mu}_{\rm min}=0.38$ and  $\delta\bar{\mu}=0.05$, compatible with the analytical estimate $\bar{\mu}=2/5$.
({\it Parameters}:  $E=7$, $\mu=-12.6$, $K=5$, $a=0$. Duration $10^6\,{\rm latt}$, lattice size $40\times40$, whole complex as initial state.)
\label{fig:scaling_mu}}
\end{figure}

\begin{itemize}
\item {\bf Determining $\mu$.} We considered relatively high values of the heterogeneity, $h\in[0.4,1.0]$, and used as scaling hypothesis ${\rm Err}/{\rm Err}_{\rm sat}=f(h/{h}_*)=g(h M^{\mu})$. Fig.~\ref{fig:scaling_mu}A shows the data before the collapse, and Fig.~\ref{fig:scaling_mu}B shows the quality function, which has a minimum at $\mu_{\rm min}=0.28$ with $\delta\mu=0.06$. This compares favorably with the theoretical prediction,  $\mu=1/3$. In Fig.~\ref{fig:scaling_mu}C we can see that for the numerically determined exponent there is a good collapse of the data. This collapse is better than that shown in Fig.~3B and C, which was done for the limiting case $K=1$. One reason for this is that in Fig.~3C the collapse was done with the analytically predicted exponent, not with the numerically determined one. Another contributing factor is that for $K=1$ the promiscuity is smaller, which can blur the transition.

\item {\bf Determining $\bar{\mu}$.} We now considered the same class of simulations as in the previous paragraph, but in the range $h\in[0.0,1.0]$ and use the scaling hypothesis $(1-{\rm Err})/(1-{\rm Err}_{\rm sat})=f(h/{h}_*)=g(h M^{\bar{\mu}})$. The calculated exponent was $\bar{\mu}_{\rm min}=0.38$, with $\delta\bar{\mu}=0.05$. This compares well with the theoretical prediction, $\bar{\mu}=2/5$. As before, Figs.~\ref{fig:scaling_mu}D to F show the raw data, the fit quality function, and the data collapse, respectively.
\end{itemize}

In the third class of simulations we fixed the number of complexes to $K=100$ and the heterogeneity to $h=1$, i.e. $M=N$. We then changed the sparsity in order to determine $\gamma$ and $\bar{\gamma}$. It was convenient to perform the analysis using $1-a$ and not $a$, as this is the quantity for which the scaling exponents were predicted. 

\begin{figure}
\centering
\includegraphics{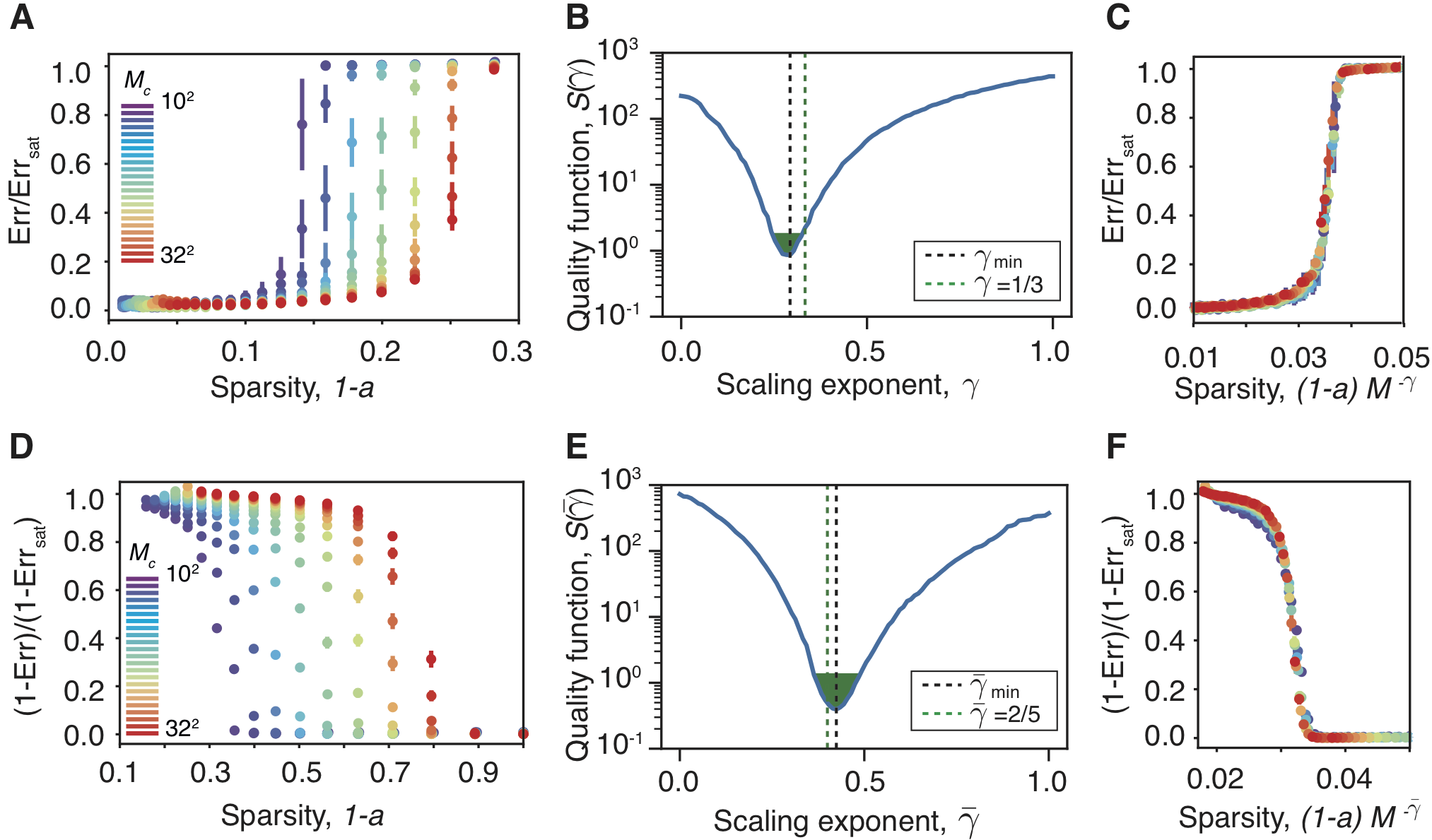}
\caption{
{\bf Scaling analysis of the sparsity, $a$.}
{\bf A.} Assembly error as a function of the sparsity. Each point is an average of $60$ simulations, with bars denoting standard deviations. At high sparsity there is a transition from MA into Ch, and the error saturates to ${\rm Err}_{\rm sat}$ (see Eq.~\ref{eq:errsat}). Different colors correspond to different sizes, see legend.
{\bf B.} The data is rescaled with different size scaling exponents, ${\gamma}$. For each exponent the quality of the collapse is quantified by the function $S(\gamma)$ as defined in \cite{houdayer2004low}, see also text. Optimal data collapse occurs at the minimum, $\gamma_{\rm min}=0.29$. We estimate $\delta\mu=0.03$ as the width of the region in which $S(\mu)$ increases by unity, shaded in green. This is in agreement with the analytical estimate $\gamma=1/3$.
{\bf C.} Collapse of the data from A using $\gamma_{\rm min}$. {\bf D}, {\bf E} and {\bf F.}  Same as A to C, but for the Ch to L transition, with $\bar{\gamma}_{\rm min}=0.42$ and  $\delta\bar{\gamma}=0.06$, compatible with the analytical estimate $\bar{\gamma}=2/5$.
({\it Parameters}:  $E=7$, $\mu=-12.6$, $K=100$, $h=1$. Duration $10^6\,{\rm latt}$, lattice size $40\times40$, whole complex as initial state.)
\label{fig:scaling_gamma}}
\end{figure}

\begin{itemize}
\item {\bf Determining $\gamma$.} We considered relatively high values of the sparsity, so that $1-a\in[0.0,0.3]$, which captures the transition from Ch to MA, see Fig.~3A. We used as scaling hypothesis ${\rm Err}/{\rm Err}_{\rm sat}=f((1-a)/(1-{a}_*))=g((1-a) M^{-\gamma})$. Fig.~\ref{fig:scaling_gamma}A shows the data before the collapse, and Fig.~\ref{fig:scaling_gamma}B shows the quality function, which has a minimum at $\gamma_{\rm min}=0.29$ with $\delta\gamma=0.03$. This compares favorably with the theoretical prediction,  $\gamma=1/3$. In Fig.~\ref{fig:scaling_gamma}C we can see that for the numerically determined exponent there is a good collapse of the data

\item {\bf Determining $\bar{\gamma}$.} We now considered the same class of simulations as above, but in the range $1-a\in[0.1,1.0]$ and using the different scaling hypothesis $(1-{\rm Err})/(1-{\rm Err}_{\rm sat})=f((1-a)/(1-{a}_*))=g((1-a) M^{-\bar{\gamma}})$. The calculated exponent was $\bar{\gamma}=0.42$, with $\delta\bar{\gamma}=0.06$. This compares well with the theoretical prediction, $\bar{\gamma}=2/5$. As before, Figs.~\ref{fig:scaling_gamma}D to F show the raw data, the fit quality function, and the data collapse, respectively.
\end{itemize}

\subsection{Boundaries between regimes in $(h, K)$ and $(a,h)$ space}
\label{sec:phase_cuts}

Fig.~3A shows a portrait in $(a, h, K)$ space of the different possible regimes for our model using Eqs.~\ref{eq:si_trans1} and \ref{eq:si_trans2}. In section \ref{sec:exponents} we verified the size scaling of $(a, h, K)$ predicted by these expressions, which is an indirect confirmation of the functional form of the boundaries that separate the different regimes. A more direct confirmation can be done by comparing numerically determined two dimensional cross-sections of this three dimensional parameter space. These cross-sections can then be compared to the predicted theoretical boundaries that separate the different regimes.

\begin{figure}
\centering
\includegraphics{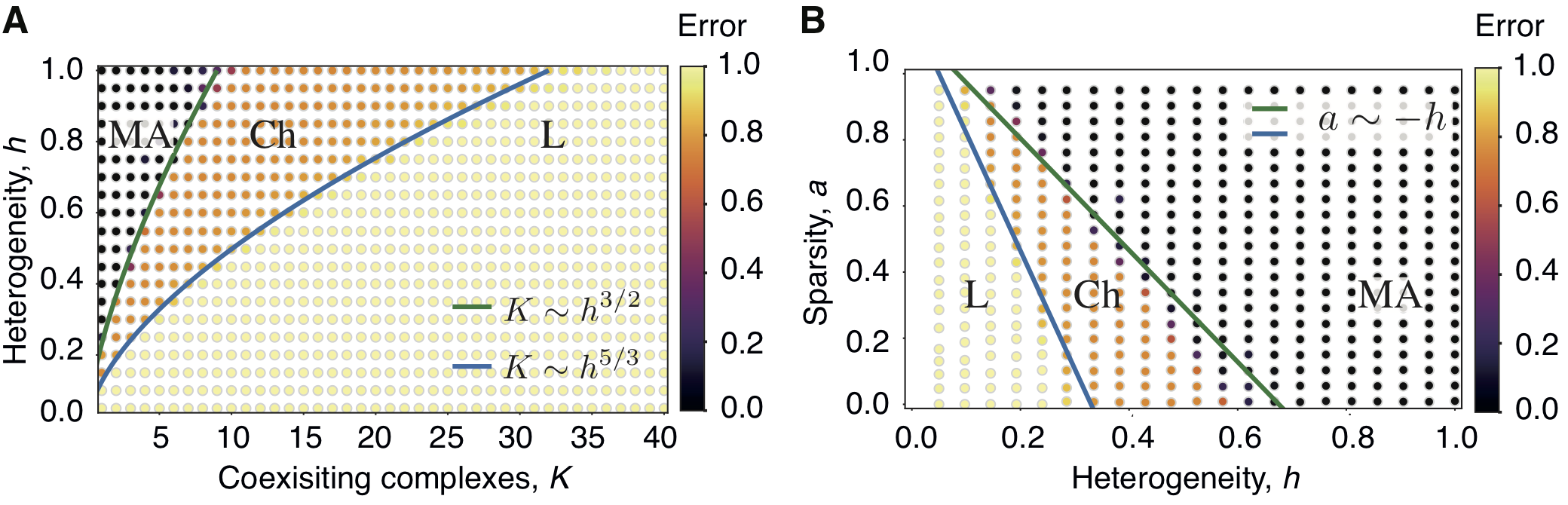}
\caption{
{\bf Orthogonal cross-sections of the parameter space.}
{\bf A} and {\bf B.} Orthogonal cross-sections of the $(a,h,K)$ space using the assembly error as order parameter. In A, $a=0$; and in B, $K=5$. The boundaries that separate the regimes have the functional forms predicted by Eqs.~\ref{eq:si_trans1} and \ref{eq:si_trans2}, see colored lines.
({\it Parameters:} $E=7$, $\mu=-12.6$, $M=20^2$. In A, $N=N_{\rm tot}$; and in B, $K=5$. Duration $10^6\,{\rm latt}$, lattice size $40\times40$, whole complex as initial state.)
\label{fig:sections}}
\end{figure}

In Fig.~\ref{fig:sections} we show two such (orthogonal) sections,  $a=0$ in A , and $K=8$ in B. A third section, $h=1$, appears in Fig.~3D.  All three cross-sections  show boundaries that match well the scaling behaviors predicted by Eqs.~\ref{eq:si_trans1} and \ref{eq:si_trans2}, and schematically depicted in Fig.~3A. Note that in order to determine the boundaries that separate the different regimes we needed to determine the exact pre-factors for the scaling relations. In Fig.~3D this was done by fitting sigmoidal functions on the $a=0$ line in the same way as described before for the boundaries in $(E,\mu)$ space. The values of $K$ obtained for the transition from MA  into Ch and from Ch into L were $9.2\pm0.1$ and $32.6\pm0.3$, respectively. These same values were then used to characterize the transition boundaries in Fig.~\ref{fig:sections}A. In the case of Fig.~\ref{fig:sections}B, besides the aforementioned values, we also used sigmoidal fits to characterize the offset of the linear boundaries. We found that, for a sparsity equal to $0.5$ the transitions from L to Ch and from Ch to MA occur at $h=0.19$ and $h=0.38$, respectively. 

\subsection{Variability in $E$ and $\mu$}
\label{si:var}
In the main text of this paper we have taken all specific interactions between components to have the same strength, $E$, and the chemical potentials of all species also to be same, $\mu$. This is clearly a simplifying assumption, as in the cellular environment we expect a large variability in these parameters. To explore the role of such a variability, we sampled the chemical potential of each species, $\mu_\alpha$, and the energy of interaction between two component species that are neighbors in a given complex, $U_{\alpha\beta}$, from two normal distributions, $\mathcal{N}(\bar{\mu}, \sigma_\mu^2)$ and $\mathcal{N}(\bar{E}, \sigma_E^2)$, respectively. We first fixed all the parameters to values such that the system finds itself in the MA regime, and systematically varied the values of $\sigma_\mu$ and $\sigma_E$. 

\begin{figure}
\centering
\includegraphics[width=\textwidth]{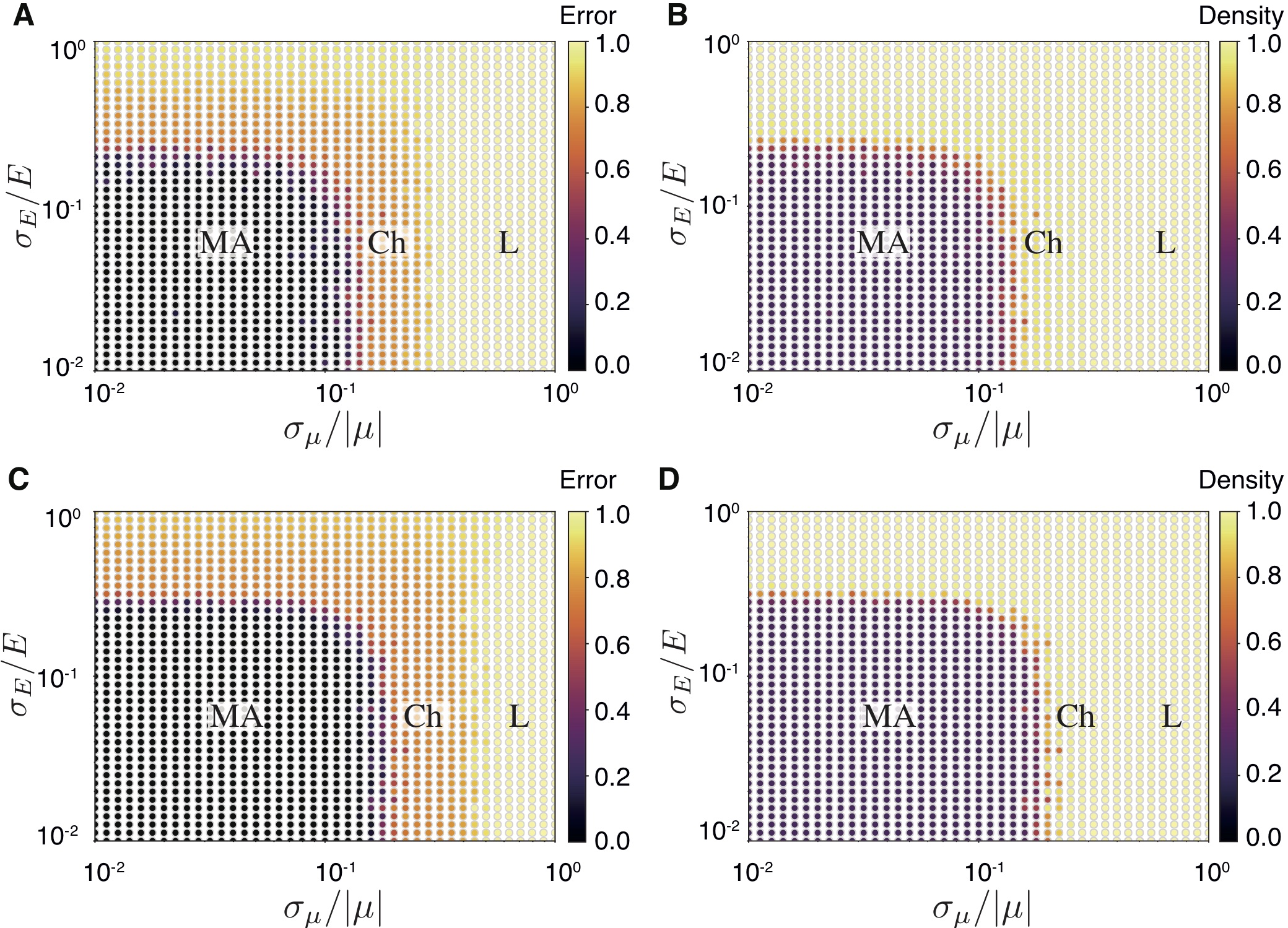}
\caption{
{\bf Effect of energy and chemical potential variability in the MA regime.}
{\bf A} and {\bf B.} Different assembly regimes as function of the standard deviation of the chemical potential, $\sigma_\mu$, and of the binding energy, $\sigma_E$, of component species. Both $E$ and $\mu$ were sampled from normal distributions, $\mathcal{N}(\bar{\mu}, \sigma_\mu^2)$ and $\mathcal{N}(\bar{E}, \sigma_E^2)$, with the means, $\bar{\mu}$ and $\bar{E}$, set to values that correspond to MA. For large enough standard deviations, the system transitions into the chimeric regime.
{\bf C} and {\bf D.} The same as A and B, but for values of $\bar{\mu}$ and $\bar{E}$ ``deeper'' into the MA regime. As a result, the robustness of MA is increased, and larger values of the standard deviations are needed to transition into Ch. 
({\it Parameters:} A and B,  $\bar{E}=12.6$, $\bar{\mu}=-7$, $K=5$. C and D, $\bar{E}=25$, $\bar{\mu}=-40$, $K=5$. The rest as in Fig.~2)
\label{fig:noise_emu}}
\end{figure}

As one can see in  Fig.~\ref{fig:noise_emu}, for very low values of the variability, the system is in the MA regime, as expected. Reliable assembly remains robust for values of the energy standard deviation up to $\sigma_E\approx0.1\bar{E}$, and values of the chemical potential standard deviation up to $\sigma_\mu\approx0.1\bar{\mu}$. Beyond this, the variability induces a transition into the chimeric regime, but not into the dilute solution regime, which also borders MA, see Fig.~2. Note that in panels C and D  in Fig.~\ref{fig:noise_emu} the transition into Ch occurs for larger values of the variability, so that the robustness of MA is increased. The reason for this is that in C and D the mean value of the energy (see caption) is larger, i.e. one is ``deeper'' in the MA regime, see Fig.~2E. Therefore, the same relative variation as in panels A and B, Fig.~\ref{fig:noise_emu}, will now result in the values of  energy for which MA is stable. We leave open the systematic study of these variability induced transitions, as well as their possible  role for the cellular environment. 

\subsection{Variability in $M_c$} \label{si:size}
In the main text of the paper we have considered the case in which all complexes have the same size, $M= M_c$. We now discuss the effect of size variability in the MA regime, i.e. its effect on the reliability of assembly. Importantly, to isolate the role of size variability in MA we have to constrain to fixed values the intensive properties of complexes, such as $h_c$ and $a_c$. Keeping this in mind, we  expect that size variability has a minor effect on MA. The reason is that the constraint of Eq.~\ref{eq:si_trans1} on MA arises from the promiscuity of components, Eq.~\ref{eq:prom}, and in terms of promiscuity it makes no difference (up to boundary effects) whether components are distributed in one large complex, or in several small ones. Therefore, as long as the average complex size is large enough to disregard boundary effects, we expect our results to be robust.  To see this quantitatively, consider the heterogeneity of complexes fixed at $h_c=1$, without any loss of generality. Following \ref{si:scale}, the promiscuity of components can be then estimated as $\pi\approx \sum M_c/N_{\rm tot}=K\langle M_c\rangle/N_{\rm tot}$, which suggests that our conclusions are not affected by the size variability.

\begin{figure}
\centering
\includegraphics{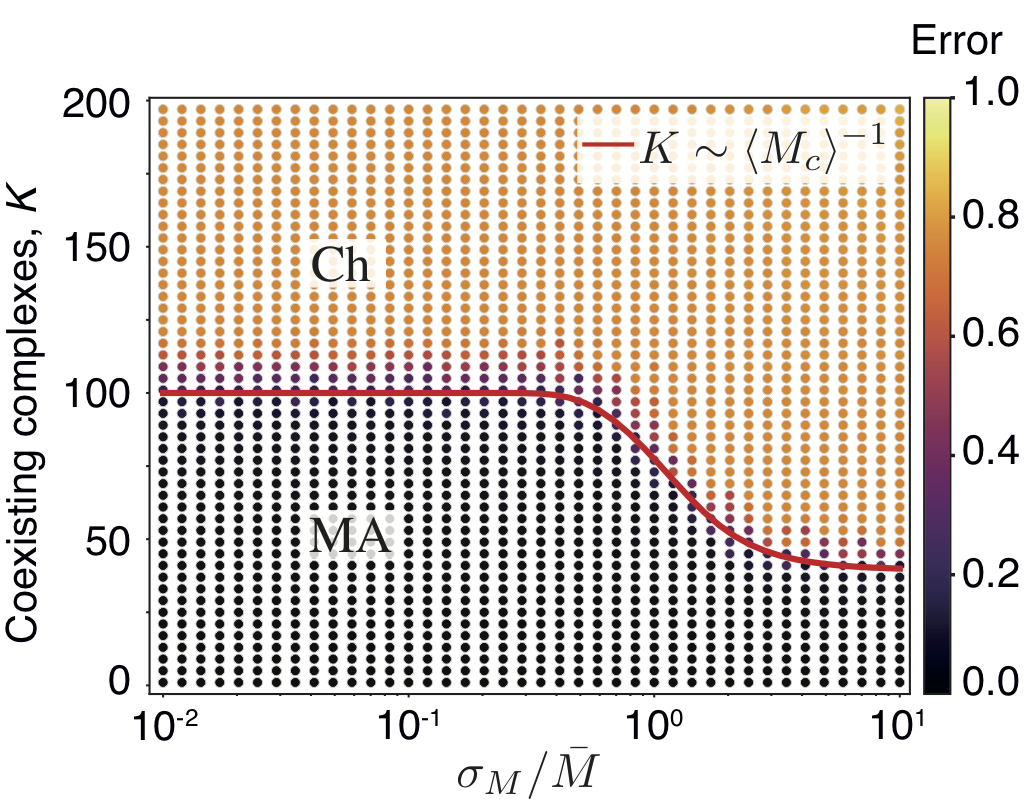}
\caption{
{\bf Effect of complex size variability in the MA regime.}
Regime diagram as a function of the number of coexisting complexes, $K$, and the variability in complex size. The size of complexes was sampled from a normal distribution, ${\mathcal N}(\bar{M},\sigma_M^2)$, truncated so that $M_c<N_{\rm tot}$. The MA regime is robust to variability in the sizes of complexes when $\sigma_M\ll\bar{M}$. When $\sigma_M\approx\bar{M}$, the truncation of the normal distribution results in a shift of the mean complex size $\langle M_c\rangle$ beyond $\bar{M}$. As a result the MA regime shrinks. The red line corresponds to the scaling predicted by the theory, taking the empirical mean of the sizes of complexes. 
({\it Parameters:} $E=7$, $\mu=-12.6$, $\bar{M}=20^2$, $N_{\rm tot}=45^2$, which gives $\langle a_c\rangle\approx0.8$ for low $\sigma_M$.   Duration $10^6\,{\rm latt}$. Lattice size $40\times40$. Whole complex as initial state.)
\label{fig:noise_M}}
\end{figure}

In order to confirm this conclusion, we performed numerical simulations, see Fig.~\ref{fig:noise_M}. In particular, we sampled the size of complexes, $M_c$,  from a normal distribution with  mean $\bar{M}$ and different values of the variance $\sigma^2_M$. We constrained the heterogeneity to $h_c=1$ by keeping $N_c=M_c$, and truncated the distribution to values $M_c<N_{\rm tot}$. Truncating the normal distribution affects the average size of complexes for large values of the variance: while for $\sigma_M\ll \bar{M}$ we have $\langle M_c\rangle = \bar{M}$, for $\sigma_M\gg \bar{M}$ we have $\langle M_c\rangle = N_{\rm tot}/2$. This increase in average complex size for large values of the variance  will affect promiscuity, and therefore alter the MA regime. Therefore, we expect that for $\sigma_M\lesssim \bar{M}$ the MA regime remains robust, as can indeed be seen in Fig.~\ref{fig:noise_M}. Conversely, for $\sigma_M\gtrsim M$ we expect that the MA regime is shrunk, as can be also observed in Fig.~\ref{fig:noise_M}. Following \ref{si:scale} it is straightforward to see that $K\lesssim \langle M_c\rangle^{-1}$, which corresponds to the red line in Fig.~\ref{fig:noise_M} ($\langle M_c\rangle$ was here empirically determined for each value of $\sigma_M$). This confirms that the shrinkage of the  MA regime is purely due to a change in the average size of complexes. Therefore, our results are robust to size variability, and they should remain valid provided that one simply replaces $M$ by $\langle M_c\rangle$.

\section*{Database analysis}
\label{sec:disclaimer}

To quantify the characteristics of complexes we analyzed publicly available databases. Broadly speaking, at the present time there are three classes of such databases for protein complexes: manually curated databases, in which the inclusion of new protein complexes is done by curators according to some pre-established criteria \cite{meldal2014complex, pu2008up}; databases generated by clustering of protein interaction networks \cite{nepusz2012detecting}, themselves generated through some pre-established experimental protocol, such as TAP-MS \cite{gavin2006proteome, krogan2006global, wan2015panorama}; and databases obtained through meta-analysis of several high-throughput datasets \cite{hart2007high,drew2017integration}. Although protein complex databases have been generated for more than a decade, they still show important weaknesses. On the one hand, manually curated databases are not comprehensive, the inclusion criteria suffer from some arbitrariness, and their growth is biased by the interests of the scientific community. On the other hand, clustering of protein interaction networks depends on the clustering algorithm, and the interaction networks themselves are known to commonly include false-positive and false-negative interactions. As far as possible, we were able to verify the conclusions of our model using several databases. However, the intrinsic limitations of protein complex databases should be kept in mind, and the conclusions derived from these databases should be taken with all necessary precautions. 

\begin{center}
\begin{table}
\begin{tabular}{|c| c|c| c| c| c| c| c|}
\hline
dataset & source & method & organism & stoichiometry & symmetry & $K$ & $N_{\rm tot}$ \\ [0.5ex] 
\hline\hline
$I$ & \cite{meldal2014complex} & manual  & {\it S. cerevisiae} & Yes & No & $238$ & $746$ \\  [1ex] 
\hline
$II$ & \cite{meldal2014complex} & manual  &  {\it S. cerevisiae}  & Yes & Yes & $59$ & $173$ \\  [1ex] 
\hline
$III$ & \cite{meldal2014complex} & manual  &  {\it S. cerevisiae}  & No & No & $589$ & $1895$ \\ [1ex] 
\hline
$IV$ & \cite{krogan2006global} & high-througput  &  {\it S. cerevisiae}  & No & No & $664$ & $2133$ \\ [1ex] 
\hline
$V$ & \cite{hart2007high} & meta-analysis  &  {\it S. cerevisiae}  & No & No & $518$ & $3457$ \\ [1ex] 
\hline
$VI$ & \cite{giurgiu2018corum}  & manual &  Multi-organism  & No & No & $2652$ & $4625$ \\ [1ex] 
\hline
$VII$ & \cite{wan2015panorama}  & high-throughput   &  Multi-organism  & No & No & $982$ & $2154$ \\ [1ex] 
\hline
$VIII$ & \cite{drew2017integration}  & meta-analysis   &  Human  & No & No & $4659$ & $7778$ \\ [1ex] 
\hline
\end{tabular}
\caption{{\bf Summary of datasets used in this work.} Datasets $I-V$ involved some degree of post-processing, as detailed in the text. For dataset $V$ we only used the ``core'' dataset, which excludes redundancies. As described in \cite{wan2015panorama}, dataset $VII$ is based on human cell data, but interactions among proteins were added provided they were confirmed for at least two other species among the five species studied.} \label{tab:datasets}
\end{table}
\end{center}

\subsection{Heterogeneity}
\label{sec:het_dat_analysis}
In order to quantify the heterogeneity  of protein complexes, we used a database with stoichiometric information. To our knowledge, the most comprehensive databases with this type of information  are those in the manually curated  Complex Portal \cite{meldal2014complex, meldal2018complex}. In particular, the database of {\it Saccharomyces cerevisiae}, which was used to generate Fig.~4, is the most complete, containing over 500 complexes. Unfortunately, not all these complexes contain full stoichiometric information. In particular, for $301$ complexes stoichiometric information is completely absent, for $50$ it is only partial (i.e. it is known only for some of its components), however for $238$ complexes in this dataset there exists  full stoichiometric information. In order not to create systematic biases in the dataset, we constrained ourselves to this last set of complexes with full stoichiometric information. Ultimately, we identified $K=238$ complexes with complete stoichiometry formed by $N_{\rm tot}=746$ components (we excluded small molecules, annotated using the  ``ChEBI'' label, such as Mg$^{2+}$), which conforms dataset $I$. The characteristics of this and other datasets are summarized in Table~\ref{tab:datasets}.

\begin{figure}
\centering
\includegraphics{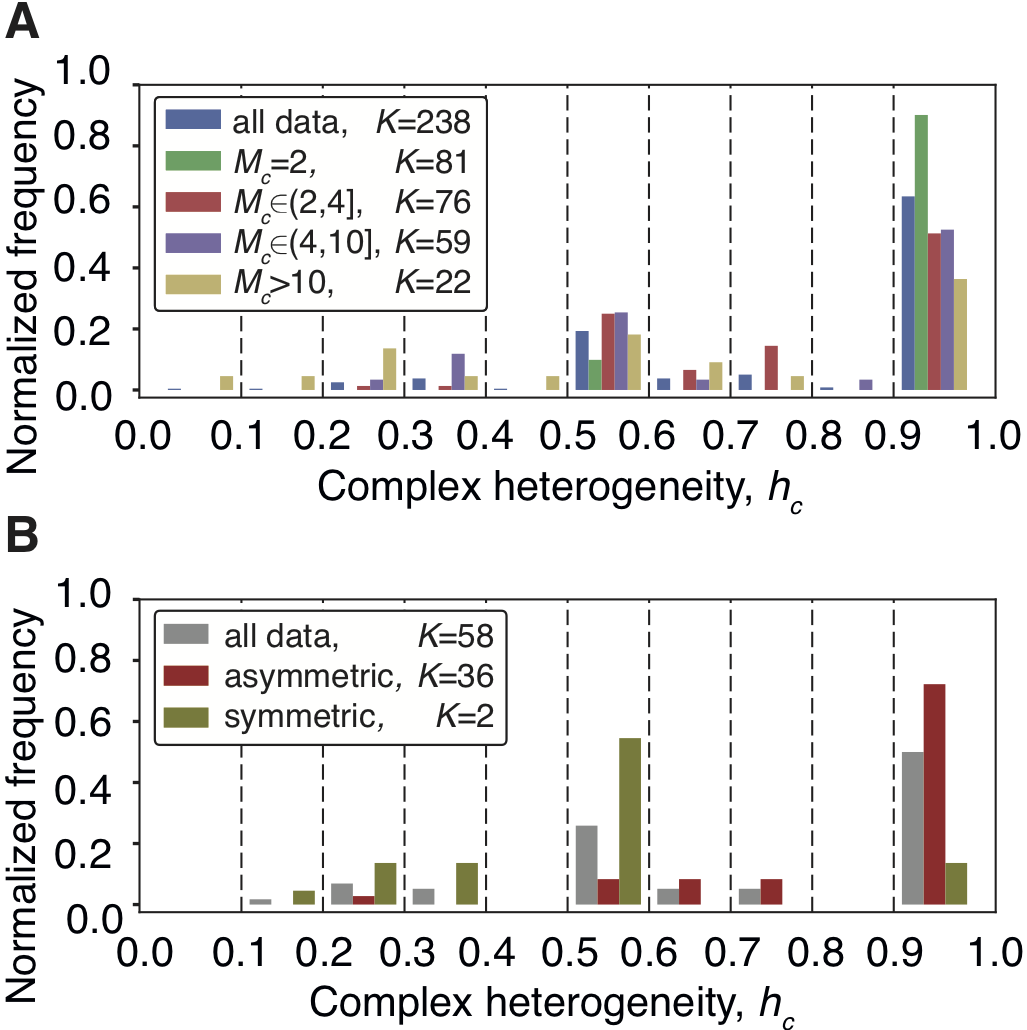}
\caption{
{\bf Heterogeneity of symmetric and asymmetric complexes.}
{\bf A.} Heterogeneity histogram for complexes in dataset $I$. Most complexes are highly heterogeneous, but a peak is also present for intermediate values of the heterogeneity. This distribution is preserved across different size ranges of complexes (different colors, see legend).
{\bf B.} Heterogeneity histogram for complexes in dataset $II$, separated into asymmetric or symmetric structures. The high heterogeneity peak is only present among asymmetric complexes, and the peak at intermediate heterogeneity only among symmetric ones.
({\it Parameters:} Bin size is half of that used in Fig.~4. Bins are within dashed lines, closed at the left and open at the right, except for $[0.9,1.0]$).
\label{fig:het_bins}}
\end{figure}

In Fig.~\ref{fig:het_bins}A we plot histograms for the complex heterogeneity using a bin size half of that used in Fig.~4A. As one can see, there is a large abundance of high-heterogeneity complexes, $h_c>0.9$, which confirms the observation of Fig.~4. The distribution is dominated by fully heterogeneous complexes, $h_c=1$, of which there are $149$. Low heterogeneity complexes, $h_c<0.2$, are rare. An example of a low heterogeneity complex is the fatty acid synthase, a dodecamer with only two species (CPX-1162, PDB: 2UV8). Importantly,  there is a persistent presence of complexes with intermediate heterogeneities, $h_c\in[0.5,0.6)$. These are dominated by complexes with heterogeneity $h_c=1/2$, of which there are $43$. For example, the Gal3-Gal80 transcription regulation complex, with two copies of each of its two components (CPX-1042, PDB: 3V2U).

To determine whether these mid-heterogeneity complexes have a relationship with  geometrical symmetries in complexes, we cross-referenced dataset $I$ with the protein data bank \cite{berman2000protein}.  We repaired some inconsistencies within the database, such as outdated PDB entries or entries referering to molecular dynamics simulations (these changes were discussed with Complex Portal curators, and should be included in future releases). The result was dataset $II$. For each complex in this dataset we looked at the symmetry group provided by the PDB website, which is calculated using alignment of repeated proteins within the complex,  as described in \url{https://www.rcsb.org/pages/help/advancedsearch/pointGroup}. We then separated the complexes in two sets, those that are asymmetric and those that exhibits a certain symmetry. In Fig.~\ref{fig:het_bins}B we plot a histogram of the heterogeneity of asymmetric and symmetric complexes for a smaller bin size than in Fig.~4D. As one can see, asymmetric complexes are much more heterogeneous than symmetric ones. In particular, the large peak in $h_c\in[0.5,0.6)$ only occurs for symmetric complexes.

\subsection{Protein participation}\label{sec:particip}

\begin{figure}
\includegraphics{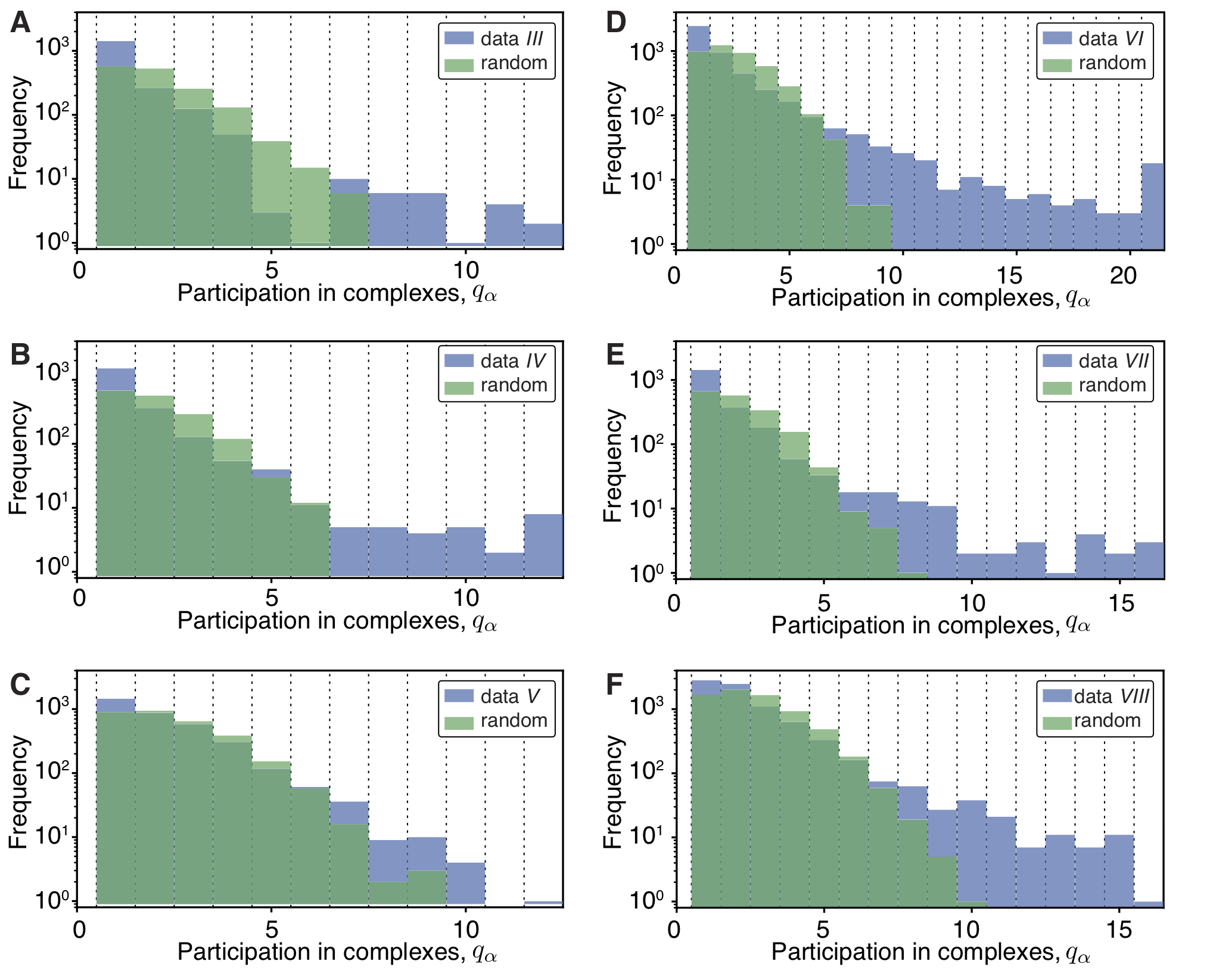}
\caption{
{\bf Protein participation is not explained by null model.}
{\bf A-F.}
Histograms of protein participation for six different datasets of protein complexes, described in Table~\ref{tab:datasets}. The null ``random model'' corresponds to randomly grouping proteins into sets (artificial imaginary complexes) with the same complex compositions as the empirically observed ones. In panels A, B, D, E, or F this null model does not account for highly participatory proteins. Only in C the null ``random model'' seems to account for the real data. Histogram bins are within dashed lines, except for the last bin, which is open ended (it includes all values higher and equal to the corresponding one).
\label{fig:random_test}}
\end{figure}

We defined the {\it participation} of a protein species $\alpha$, $q_
\alpha$, as the number of complexes in which $\alpha$ takes part. Importantly, to determine this quantity the only information about complexes needed was their composition (no stoichiometric or structural information is relevant here). We thus defined dataset $III$, larger than $I$ or $II$, as the raw Complex Portal dataset (excluding, again, small molecules). For this dataset $III$ we quantified $q_\alpha$ through a histogram, Fig.~\ref{fig:random_test}A (blue), which included  protein species that participate in seven or more complexes (there are $29$ such proteins). An example of the latter is Cdc28 (P00546), the yeast cyclin dependent kinase, which participates in nine complexes. We also quantified the number of species per complex, $N_c$, which spanned about two orders of magnitude. Given that both quantities spanned a wide range of values, we asked whether a simple null model could explain the histogram of $q_\alpha$ using the histogram of $N_c$ as input. To answer this question, we generated $K$ artificial complexes by randomly sampling the $N_{\rm tot}$ different protein species into sets with $N_c$ different species, where $N_c$ was taken from the original dataset. In this way, the resulting ``random'' dataset was guaranteed to generate the same histogram of $N_c$. However, as shown in Fig.~\ref{fig:random_test}A (green), the histogram of $q_\alpha$ for the random dataset deviated from that of dataset $III$, not recovering the high participation of some proteins.

If confirmed, this is a somewhat surprising conclusion in view of the recent observation in Ref.~\cite{mazzolini2018statistics} implying that for several complex systems the  distribution of $q_\alpha$ (therein called ``occurence'', up to normalization) could be explained by a random sampling of components into sets of sizes given by the empirically observed compositions (therein including stoichiometry). Our procedure is analogous to that of Ref.~\cite{mazzolini2018statistics}, the only difference being that we disregarded stoichiometry. Our initial expectation was that the random sampling approach would recover the distribution of $q_\alpha$. However, the disparity depicted in Fig.~\ref{fig:random_test}A suggests  that the participation of proteins cannot be explained by randomly sampling the composition of complexes. 

We sought to confirm this conclusion by analyzing alternative datasets. Once the need for stoichiometric information is lifted, there are several available databases of protein complexes. We used two additional datasets for {\it S. cerevisiae}. To remove the possibility that our conclusions are an artifact of manual curation, we used the high-throughput data from \cite{krogan2006global} to generate dataset $IV$. In particular, we applied the recent ClusterOne clustering algorithm from \cite{nepusz2012detecting} to the data from \cite{krogan2006global} in order to generate  dataset $IV$. ClusterOne has been shown to outperform other standard clustering algorithms (note also that the commonly used MCL  algorithm \cite{enright2002efficient} does not allow for complexes to share components). In particular, we used the weighted version of the algorithm, setting the probability of interaction as weight (the parameters of the algorithm were the size and threshold of the detected complexes, taken to be $2$ and $0.3$, respectively). To generate  dataset $V$ we used the meta-data from \cite{hart2007high}, which combines the raw data from \cite{krogan2006global} and \cite{gavin2006proteome} and assigns a $p-$value for the interactions. Again, we clustered the data using the ClusterOne algorithm with weights given by $1-p$ (parameters as before). The histograms of $q_\alpha$ for datasets $IV$ and $V$ are shown in panels B and C of Fig.~\ref{fig:random_test}. Panel B exhibits a very similar trend to panel A, with highly participatory proteins that cannot be explained by our simple null model. Panel C, however, is compatible with the random model. 

To further support the observation of a broad distribution of $q_\alpha$ that cannot be explained by the random model, we sought larger datasets, with more protein species and more complexes. Datasets $VI$, $VII$, and $VIII$ include metazoan data, and are all significantly larger than the yeast datasets. Dataset $VI$ is the manually curated Corum 3.0 dataset \cite{giurgiu2018corum}, which incorporates data of humans ($67\%$), mouse ($15\%$), and rat ($10\%$). Specifically, it corresponds to the ``core'' dataset, in which cross-species redundancies are removed (this was the dataset used in Fig.~5). Dataset $VII$ corresponds to the high-throughput experiments carried out and analyzed in \cite{wan2015panorama}. Here, the data was clustered into complexes using two steps: first, a machine learning classifier trained with part of the Corum dataset; second, the ClusterOne algorithm. Finally, dataset $VIII$ corresponds to the meta-analysis of several human high-throughput experiments \cite{drew2017integration}, with data from \cite{wan2015panorama} as well as \cite{hein2015human} and \cite{huttlin2015bioplex}. The data was again classified using a two-step procedure similar to the one used for $VII$. The histograms for datasets $VI-VIII$ are shown in panels D, E and F of Fig.~\ref{fig:random_test}. In all these cases the distribution of the protein participation is broad, and deviates significantly from the ``random model''. This provides further support to the hypothesis that the statistics of participation of proteins in complexes can not be derived from the statistics of protein complex composition. It remains an open question what are the causes for the observed prevalence of highly participatory proteins.

\end{appendices}
\end{document}